\documentclass[preprint,10pt,numbers]{sigplanconf}

\usepackage{amsmath}
\usepackage{xspace}
\usepackage{filecontents}
\usepackage[lined,figure]{algorithm2e}
\usepackage{graphicx}
\usepackage{caption}
\usepackage{subcaption}
\usepackage{longtable}
\usepackage{ctable}
\usepackage{booktabs}
\usepackage{url}
\usepackage{amsfonts}
\usepackage{pbox}
\usepackage{comment}
\usepackage{hyperref}

\definecolor{forestgreen(web)}{rgb}{0.13, 0.55, 0.13}

\usepackage{listings}          
\lstdefinelanguage{Julia}%
  {morekeywords={abstract,break,case,catch,const,continue,do,else,elseif,%
    end,export,false,for,function,immutable,import,importall,if,in,%
        macro,module,otherwise,quote,return,switch,true,try,type,typealias,%
        using,while,Int,Int64,Float64,Matrix,Vector},%
  ndkeywords={@acc,hpat},%
  ndkeywordstyle=\color{red},%
  otherkeywords={*,\.*,:,-,.,.==,^,=,<,>},
  sensitive=true,%
  alsoother={$},%
  morecomment=[l]\#,%
  morecomment=[n]{\#=}{=\#},%
  morestring=[s]{"}{"},%
  morestring=[m]{'}{'},%
}[keywords,comments,strings]%

\makeatletter
\newcommand{\labitem}[2]{%
\def\@itemlabel{\textbf{#1}}
\item
\def\@currentlabel{#1}\label{#2}}
\makeatother

\newcommand{\Comment}[1]{}
\newcommand{\Space}[1]{} 

\newcommand{\mpic}{{MPI/C++}\xspace}
\newcommand{\mpiompc}{{MPI/OpenMP/C++}\xspace}
\newcommand{\cpp}{{C++}\xspace}
\newcommand{\charm}{{Charm++}\xspace}

\newcommand{\spark}{Spark\xspace}

\definecolor{comment-red}{rgb}{0.8,0,0}

\newcommand{\INTELR}{Intel\textsuperscript{\textregistered}}
\newcommand{\XEONR}{Xeon\textsuperscript{\textregistered}}

\newtheorem{heuristic}{Heuristic}

\usepackage{lstlinebgrd} 

\begin{document}


\title{HPAT: High Performance Analytics with Scripting Ease-of-Use}

\authorinfo{Ehsan Totoni}
{Intel Labs, USA}
{ehsan.totoni@intel.com}
\authorinfo{Todd A. Anderson}
{Intel Labs, USA}
{todd.a.anderson@intel.com}
\authorinfo{Tatiana Shpeisman}
{Intel Labs, USA}
{tatiana.shpeisman@intel.com}

\maketitle

\begin{abstract}
Big data analytics requires high programmer productivity and high performance
simultaneously on large-scale clusters. However, current big data analytics frameworks
(e.g. Apache Spark) have prohibitive runtime overheads since they are library-based.
We introduce a novel auto-parallelizing compiler approach
that exploits the characteristics of the data analytics domain
such as the map/reduce parallel pattern and is robust,
unlike previous auto-parallelization methods.
Using this approach, we build High Performance Analytics Toolkit (HPAT),
 which parallelizes high-level scripting (Julia) programs automatically,
generates efficient MPI/C++ code, and provides resiliency.
Furthermore, it provides automatic optimizations for scripting programs, such as fusion of
array operations.
Thus, HPAT is 369$\times$ to 2033$\times$ faster than Spark on the Cori
supercomputer and 20$\times$ to 256$\times$ times on Amazon AWS.



\end{abstract}

%

 \keywords{Big Data Analytics, High Performance Computing,
  Automatic Parallelization}

\section{Introduction}\label{sec:intro}
Big data analytics applies advanced analytics and machine learning
techniques to gain new insights from large data sets, which are gathered from
sources such as sensors, web, log files, and social media.
Big data analytics allows users to extract knowledge from this data
 and make better and faster decisions.
However, supporting fast decision making necessitates rapid development
of the application by domain expert programmers (i.e., high productivity)
 and low execution time (i.e., high performance).
High productivity in the data analytics domain requires
 scripting languages such as MATLAB, R, Python, and Julia
since they express mathematical operations naturally
and are the most
productive languages in practice~\cite{Prechelt:2000,Chaves06}.
High performance requires efficient execution on large-scale distributed-memory clusters
due to extreme dataset sizes. 



Currently, there is a significant productivity and performance gap in the big data analytics domain.
A typical High Performance Computing (HPC) approach of writing low-level codes (e.g. \mpic)
 is beyond the expertise of most data scientists and is not practical in their
 interactive workflows.
Existing big data analytics frameworks
such as Apache Hadoop~\cite{white2012hadoop} and Apache Spark~\cite{zaharia2010spark}
provide better productivity for big data analytics on clusters using the MapReduce programming paradigm~\cite{dean2008mapreduce}.
However, this productivity comes at the cost of performance as
these frameworks are orders of magnitude slower than hand-written
\mpic programs~\cite{brown2016have,ReyesOrtizMPIspark2015,mcsherry2015scalability}.
A fundamental issue is that these frameworks are library-based,
requiring a runtime system to coordinate parallel execution across all the nodes.
This leads to high runtime overheads - for example, the master
node managing execution is typically a bottleneck.


We propose a novel auto-parallelizing compiler approach for this domain that
does not suffer from the shortcomings of prior methods~\cite{Wilson:1994:SIR,Blume:1996:PPP,Kennedy:1998:ADL,Chatterjee:1995:OEA}.
These efforts failed in practice due to the complexity of the problem,
especially for distributed-memory architectures~\cite{Eigenmann94onthe,Waheed99cap,Hiranandani:1994:DEN,Kennedy:2007:RFH}.
The main challenge is that the compiler needs to know which data structures and loops of the program should be parallelized,
 and how they should be mapped across the machine.
Previous algorithms for automatic parallelization typically start by
analyzing array accesses of the program and finding all possible array and computation
distributions for loop-nests. Then, distributions are selected based on heuristics and/or cost models.
A fundamental problem is that the approximate cost models and heuristics
cannot find the best combination of distributions from a potentially large search space reliably,
leading to significantly lower performance compared to manual parallelization.
In contrast, we develop a data flow algorithm that exploits domain knowledge as well as
high-level semantics of mathematical operations to find the best distributions,
 but without using approximations such as cost models.
Hence, our approach is robust and matches manual parallelization.


In this paper, we use this approach to build High Performance Analytics Toolkit (HPAT)\footnote{HPAT is available
online as open-source at \url{https://github.com/IntelLabs/HPAT.jl}.}.
HPAT is a compiler-based framework for big data analytics on large-scale clusters
that automatically parallelizes high-level analytics programs
and generates scalable and efficient \mpic code.
%
To the best of our knowledge, HPAT is the first compiler-based system that can parallelize
 analytics programs automatically and achieve high performance.
Our contributions can be summarized as follows:
\begin{itemize}
\item A robust domain-specific automatic parallelization algorithm (\textsection\ref{sec:inference})
\item Domain-specific optimization techniques (\textsection\ref{sec:domainopt})
\item A system design addressing various challenges such as parallel I/O code generation (\textsection\ref{sec:distributed})
\item A domain-specific automatic checkpointing technique (\textsection\ref{sec:checkpointing})
\end{itemize}

Our evaluation demonstrates that HPAT is 369$\times$ to 2033$\times$ faster
than Spark on the Cori supercomputer~\cite{coriNERSC}
and provides similar performance to hand-written \mpic programs.
HPAT is also 20$\times$-256$\times$ faster on Amazon AWS (\textsection\ref{sec:evaluation}).
\section{Logistic Regression Example}\label{sec:example}
\begin{figure}
\centering
\begin{subfigure}[b]{\columnwidth}
\begin{lstlisting}[language=Julia,escapeinside={(*}{*)},mathescape]
using HPAT

@acc hpat function logistic_regression(iters,file)
  points = DataSource(Matrix{Float64},HDF5,"points",file)
  labels = DataSource(Vector{Float64},HDF5,"labels",file)
  D,N = size(points)
  labels = reshape(labels,1,N)
  w = 2*rand(1,D)-1
  for i in 1:iters
    w -= ((1./(1+exp(-labels.*(w*points)))-1).*labels)*points'
  end
  return w
end

weights = logistic_regression(100,"mydata.hdf5")
\end{lstlisting}
\caption{HPAT logistic regression example (Julia).}
\label{fig:hpatlr}
\end{subfigure}
\begin{subfigure}[b]{\columnwidth}
\begin{lstlisting}[language=Python,escapeinside={(*}{*)},mathescape]
from pyspark import SparkContext

D = 10  # Number of dimensions
# read batch of points for faster Numpy computation.
def readPointBatch(iterator):
  strs = list(iterator)
  matrix = np.zeros((len(strs), D + 1))
  for i, s in enumerate(strs):
    matrix[i] = np.fromstring(s.replace(',',' '),sep=' ')
  return [matrix]

if __name__ == "__main__":
  sc = SparkContext(appName="PythonLR")
  points = sc.textFile(sys.argv[1]).mapPartitions(readPointBatch).cache()
  iterations = int(sys.argv[2])
  w = 2*np.random.ranf(size=D)-1

  def gradient(matrix, w):
    Y = matrix[:, 0]    # unpack point labels
    X = matrix[:, 1:]   # unpack point coordinates
    return ((1.0 / (1.0 + np.exp(-Y * X.dot(w))) - 1.0) * Y * X.T).sum(1)

  def add(x, y):
    x += y
    return x

  for i in range(iterations):
    w -= points.map(lambda m: gradient(m, w)).reduce(add)
  sc.stop()
\end{lstlisting}
\caption{Spark logistic regression example (Python)}\label{fig:sparklr}
\end{subfigure}
\caption{Logistic regression example codes.}
\label{fig:lrcode}
\end{figure}

We introduce HPAT and compare it to the state-of-the-art using
the logistic regression machine learning algorithm.
Logistic regression uses the {\em logistic sigmoid} function to derive
a model for data classification~\cite{Bishop:2006:PRM}. This model is updated
iteratively using the gradient descent optimization method.
Figure~\ref{fig:hpatlr} presents the HPAT code for logistic regression.
The \emph{@acc hpat} macro annotation and the \emph{DataSource} construct are HPAT-specific,
but the rest of the code is high-level sequential Julia code which uses matrix/vector
operations (corresponding to ``vectorized'' mathematical notation).
HPAT automatically parallelizes and optimizes this program and
generates efficient \mpic. HPAT's auto-parallelization algorithm
uses domain knowledge to infer distributions for all variables and
computations accurately; for example, vector {\em w}
is replicated on all processors while columns of matrix {\em points} are divided in block manner,
which is the same strategy for manual parallelization of this program.
Previous auto-parallelization methods cannot achieve this since there is
a large distribution space even for this simple program;
for example, even computations on {\em w} are data-parallel and could be distributed
without violation of dependencies, but parallel performance suffers.
HPAT also fuses all computation steps in the iteration loop
together to achieve best data locality.

The current state-of-the-art for big data analytics frameworks is
the library approach, which can have significant overheads.
We use Apache Spark to demonstrate this approach since it is a widely used big data analytics
framework -- other frameworks are similar in principle.
Figure~\ref{fig:sparklr} presents the Spark version, which initializes a
 {\em resilient distributed dataset} (RDD) called \emph{points}.
 An RDD is essentially a distributed data structure that is distributed in one dimension.
In addition, the computation is written in terms of
\emph{map} and \emph{reduce} operators. In each iteration,
the task scheduler of Spark runtime library in \emph{master} node
divides map/reduce operators into tasks, and sends these tasks to the \emph{executor} (slave) nodes.
The operator closures which include the required data (\emph{w} in this case)
are also serialized and sent.
On {\em executors}, each task is initialized and the RDD data is deserialized from the data cache into Python (Numpy) objects
for execution. Finally, the result of the \emph{reduce} operator is sent back to the \emph{master} node.
This sequence is repeated for every iteration since the library
has to return the results of reductions to the user context and is not aware of
the iteration loop in the Python code.
In essence, each iteration is a separate job that is launched as
a wave of small tasks; scheduling and serialization overheads incur repeatedly.
This is a fundamental problem with this library design and cannot be mitigated easily --
significant Spark development effort has not closed the performance gap
with \mpic codes
(\spark has over 1000 contributors~\cite{sparksurvey2016}).

\begin{figure}
    \centering
    \def\svgwidth{.9\columnwidth}
    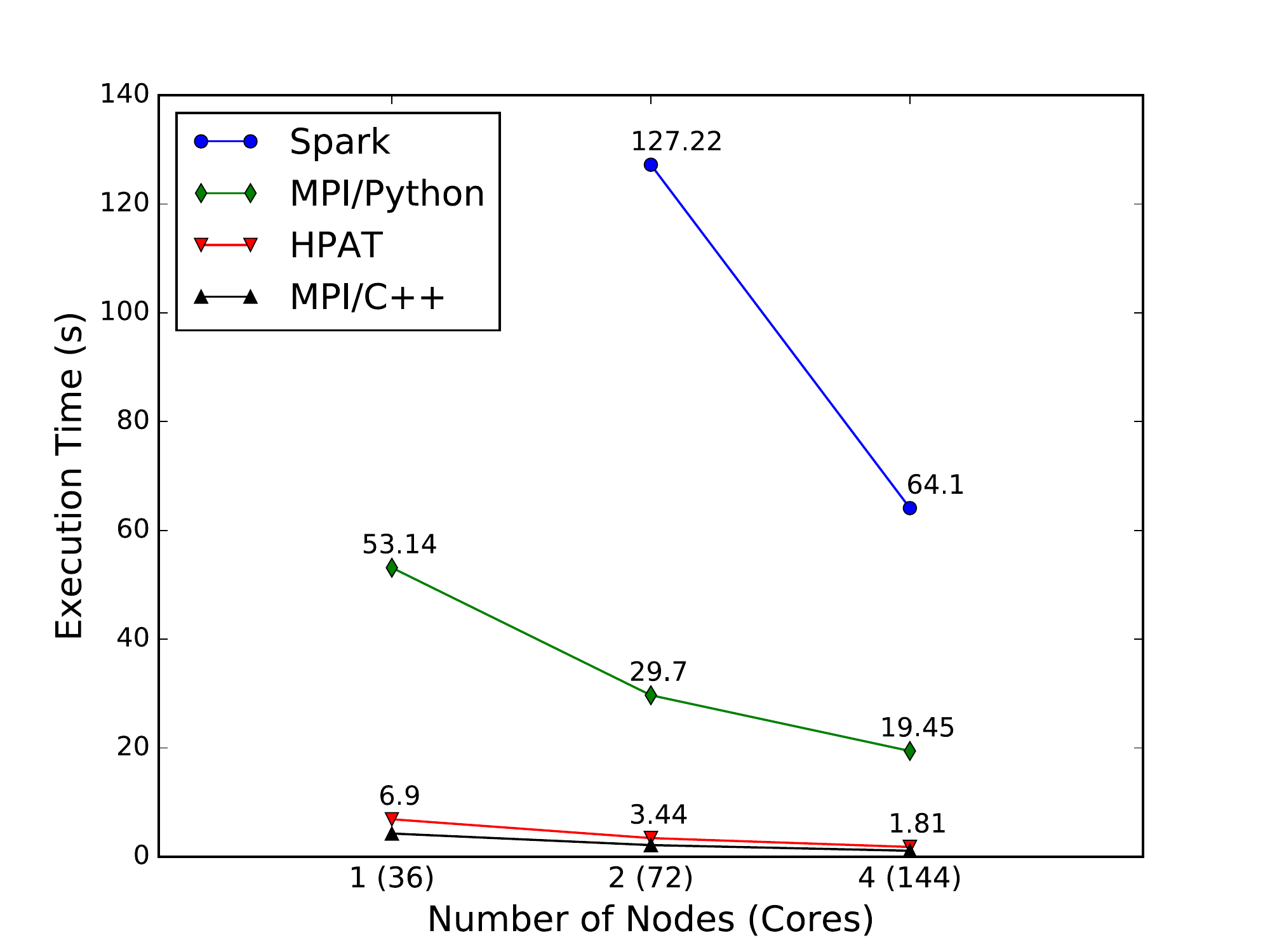
    \caption{Performance of Logistic Regression on Amazon AWS (c4.8xlarge instances, 256M 10-feature samples,
    20 iterations).}
    \label{fig:lraws}
\end{figure}

Figure~\ref{fig:lraws} demonstrates the performance of logistic regression algorithm
in Spark, MPI/Python, HPAT, and \mpic. HPAT is 35$\times$ faster than Spark and is close
to the handwritten \mpic code. The gap between Spark and the MPI/Python code
reveals the magnitude of overheads in Spark. 
The gap between MPI/Python code and \mpic is mostly due to locality advantages of
fused loops as opposed to executing a sequence of Numpy operations.
Bridging the performance gap with \mpic codes is crucial for big data analytics;
for example, 91\% of \spark users chose performance among the most important aspects for them
in a \spark survey~\cite{sparksurvey2015}. 
Furthermore, distributed libraries are implemented in languages such as Java and
Scala since reflection, serialization, and isolation features of a sandbox like Java Virtual Machine (JVM)
are required, but it
can have significant overheads~\cite{martin2015,david2016,OusterhoutAnalysis15}.
In contrast, HPAT achieves scripting productivity and HPC performance
simultaneously using an auto-parallelizing compiler approach.


\section{HPAT Overview}\label{sec:distributed}

In this section, we provide an overview of our target application domain and the HPAT compilation pipeline.

\paragraph{Domain Characteristics:}
The goal of HPAT is to automatically parallelize common analytics tasks
such as data-parallel queries and iterative machine learning algorithms.
The main domain characteristic we exploit is that the map/reduce parallel pattern
inherently underlies the target application domain.
This characteristic is assumed in current frameworks like Spark as well.
Hence, distribution of parallel vectors and matrices is mainly one-dimensional (1D) for analytics tasks
(i.e. matrices are ``tall-and-skinny'').
Furthermore, the workload is uniform across the data set (which can be extended by providing load balancing).
These assumptions do not necessarily apply to other domains. For example,
multi-dimensional distribution is often best in many physical simulation applications.
Note that while traditional HPC applications are sometimes developed over many years,
analytics programs need to be developed in as little as a few minutes in some cases to support the interactive
workflow of data scientists (which makes productivity of scripting languages critical).
Hence, analytics programs are often significantly shorter and simpler than many HPC applications.

\paragraph{HPAT Coding Style:}
HPAT supports the high-level scripting syntax of the Julia language,
which is intuitive to domain experts. However, users need to follow certain coding style guidelines
to make sure HPAT can analyze and parallelize their programs automatically:
\begin{itemize}
	\item The analytics task is in
	 functions annotated with \emph{@acc hpat}.
	\item I/O (e.g. reading input samples) is performed
	through HPAT (using the \emph{DataSource} and \emph{DataSink} constructs).
	\item The data-parallel computations are in the form of high-level matrix/vector
	 computations or comprehensions since HPAT does not parallelize sequential loops.
	\item Julia's column-major order should be followed for multidimensional arrays
	since HPAT parallelizes across last dimensions.
	For example, this means that features of a sample are in a column of the samples matrix.
\end{itemize}

\begin{figure}
\centering
\includegraphics[width=.9\linewidth]{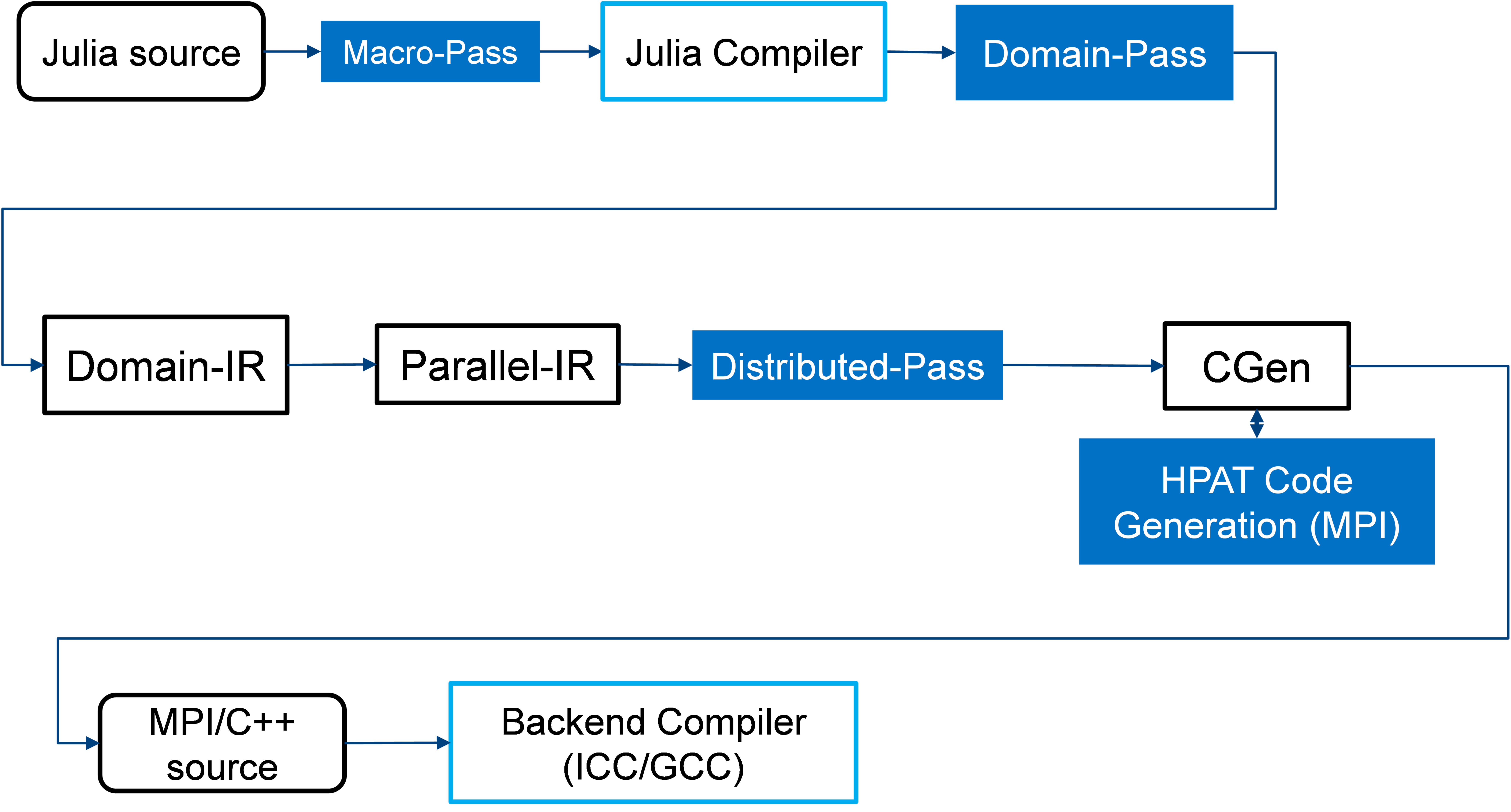}
\caption{HPAT's compilation pipeline.}
\label{fig:pipeline}
\end{figure}

\paragraph{HPAT Compiler Pipeline:}
HPAT uses the \emph{@acc} macro provided by Julia's CompilerTools package to configure HPAT's
compilation pipeline as shown in Figure~\ref{fig:pipeline}.
The solid boxes are HPAT components.
The macro pass desugars HPAT extensions (such as \emph{DataSource}) into
function calls and type annotations to enable compilation by Julia.
The Julia compiler then performs further desugaring and type inference on the function's intermediate representation (IR).
The Domain-Pass transforms HPAT extensions into a form more conducive to optimization by the subsequent Domain-IR and Parallel-IR passes.
The Domain-IR and Parallel-IR passes are provided by Julia's ParallelAccelerator package~\cite{paGithub}.
The Domain-IR pass identifies operators (e.g. vector-vector element-wise addition) and other constructs in the IR whose semantics are parallelizable.
The Parallel-IR pass takes the different kinds of parallel constructs found by Domain-IR and transforms those into a common representation
called \emph{parfor} (that represents a tightly nested \emph{for} loop whose iterations can all be performed in parallel)
and performs fusion and other optimizations on the IR.
These two passes are described in more detail in Section~\ref{sec:parallelaccelerator}.
Given this parallel representation, the Distributed-Pass partitions the arrays and \emph{parfors} for
distributed-memory execution and generates communication calls.
HPAT Code Generation takes advantage of several hooks provided by \emph{CGen} (part of ParallelAccelerator)
to generate C++ code. 

\section{Automatic Parallelization}\label{sec:inference}

Our novel auto-parallelization algorithm exploits domain knowledge
in a {\em data flow framework}~\cite{nielson2015principles} to find data and computation distributions.
The goal of the algorithm is to find a consistent map/reduce view of the program
that does not violate the high-level parallel semantics of any mathematical operation.
Since the underlying parallel pattern is map/reduce, each array should be either distributed
in 1D block fashion (\emph{1D\_B}) or replicated on all processors (\emph{REP}).
We also support a 2D block-cyclic distribution (\emph{2D\_BC}) for the less
 common square matrix computations, but this requires a user annotation.
To give some intuition, 1D block distribution is typically applied to arrays such as datasets and map operations on
those datasets, whereas a replicated distribution is used for an array
 containing a model and is often associated with a reduction.
We also know that data remapping is not needed in this domain so array distributions are global.
On the other hand, the distribution of each computation step can be determined
based on its high-level semantics (e.g. reductions) and the distribution of its inputs/outputs.

We construct our data flow framework as follows.
The \emph{meet-semilattice} of distributions
is defined as:

\centerline{$L = \{1D\_B, 2D\_BC, REP\}$, $REP \leq 2D\_BC \leq 1D\_B$}

\centerline{$\bot = REP$, $\top = 1D\_B$}

\noindent The properties to infer are defined as

\centerline{$\mathcal{P}_a: \mathbb{A} \rightarrow L, \mathcal{P}_p: \mathbb{P} \rightarrow L$}

\noindent where $\mathbb{A}$ is the set of arrays in the program,
$\mathbb{P}$ is the set of parfors,
$\mathcal{P}_a$ specifies distributions for arrays, and
$\mathcal{P}_p$ specifies distributions for parfors.
Other operations such as matrix multiply and library calls
also have similar properties, which we omit here for brevity.
Next, we define the set of {\em transfer functions} $\mathcal{F}$ for each
node type based on its parallel semantics, which are used
to update the properties.
\noindent In essence, this framework provides equation

\centerline{$ (\mathcal{P}_a,\mathcal{P}_p) = \mathcal{F}(\mathcal{P}_a,\mathcal{P}_p)$}

\noindent which can be solved
using a {\em fixed-point iteration} algorithm (repeatedly walks over the IR until quiescence).
The initial distributions are assigned as $1D\_B$
for all arrays and parfors (the top element in the semi-lattice).
We design the transfer functions to be monotone, which is consistent with the semantics of
the operations in this domain since data remappings are not common.
However, it is possible to remap data by inserting special remapping nodes that copy data to new arrays (left for future work).
Monotonicity ensures that the fixed-point iteration algorithm converges to the {\em least solution}~\cite{nielson2015principles},
which means that higher values like $1D\_B$ are preserved as much as possible while satisfying all the equations.
This ensures maximum parallelization for the program.
Program control flow (e.g. branches) can be safely ignored
 since the distributions do not change across different paths in this domain.

\noindent \emph{Transfer Function: Array Assignment - }
The distribution of left-hand side and right-hand side of an assignment on arrays
should have the same distribution, which is the {\em meet} ($\wedge$) of
their previously inferred distributions:

\centerline{$f_{l=r} : \mathcal{P}_a(l) = \mathcal{P}_a(r) = \mathcal{P}_a(l) \wedge \mathcal{P}_a(r)$}

\noindent \emph{Transfer Function: Calls - }
When the distribution algorithm encounters a call site, the algorithm determines whether it knows the transfer function for the function being called.
If so, the call site is considered to be a \emph{known call}; otherwise, it is called an \emph{unknown call}.
For unknown calls, such as functions from external modules not compiled through HPAT,
the distribution algorithm conservatively assumes the involved arrays need to be \emph{REP}.
If the function has parallel semantics for arrays, the user needs to provide the information.
Conversely, distribution transfer functions are built into a HPAT \emph{knownCallProps} table for many Julia and HPAT operations.
Common examples include Julia's array operations (e.g. reshape, array set, array get, array length),
and HPAT's data storage functions.

\centerline{$f_{unknown\ call\ g(x)} : \mathcal{P}_a(x) = REP$}

\centerline{$f_{known\ call\ g(x)} : \mathcal{P}_a(x) = \mathcal{P}_a(x) \wedge knownCallProp(g)$}

\noindent \emph{Transfer Function: Returned Array - }
For return statements of the top-level function,
 the set of arrays being returned are each flagged as \emph{REP}
 since returned arrays need to fit on a single node and
this output typically represents a summarization of a much larger data set.
If larger output is required, the program should write the output to storage.
This is a useful domain-specific heuristic that facilitates compiler analysis.

\centerline{$f_{return\ x} : \mathcal{P}_a(x) = REP$}

\noindent \emph{Transfer Function: Matrix Operations - }
Matrix/matrix and matrix/vector multiply operations (GEMM/GEMV call nodes)
have more complex transfer functions.

\centerline{$f_{lhs=gemm(x,y)} : GemmTransfer(lhs,x,y)$}

Figure~\ref{fig:gemminfer} illustrates the portion of the GEMM transfer function that is exercised during auto-parallelization of
the logistic regression example of Figure~\ref{fig:hpatlr}, as well as a \emph{2D\_BC} case. 
These formulas are derived based on our matrix distribution and layout assumptions, and semantics
of GEMM operations.
Since samples are laid out in columns and we do not split sample features across processors (\emph{1D\_B}),
any vector or row of a matrix computed using reduction across samples is inferred as \emph{REP}.
For example, the result of the inner formula in Logistic Regression's kernel is multiplied by the transpose
of sample points and used to update the parameters ($w -= (\cdots.*labels)*points'$).
Using this analysis, the algorithm infers that the output of the operation is \emph{REP} if both inputs are \emph{1D\_B}
and the second one is transposed. This also means that the inputs can stay \emph{1D\_B}.
In this case, a reduction is also inferred for the node (which eventually turns into MPI\_Allreduce in the backend).
Furthermore, since the vector \emph{w} is used in a dot product with matrix columns in $w*points$,
it should be \emph{REP} and the matrix stays as \emph{1D\_B}.

\begin{algorithm}
\DontPrintSemicolon
\SetKwFunction{procgemm}{GemmTransfer}
\SetKwFunction{isseq}{isREP}
\SetKwFunction{isod}{is1D}
\SetKwFunction{istd}{is2D}
\SetKwFunction{istran}{isTransposed}
\SetKwProg{myproc}{Procedure}{}{}
\myproc{\procgemm{lhs,x,y}}{
	\uIf{\isod{x} $\land$ \isod{y} $\land$ $\neg$\istran{x} $\land$ \istran{y}}{
	\tcp{e.g. ($\cdots$.*labels)*points' - reduction across samples}
        {$\mathcal{P}_a(lhs) = REP$}
	}\uElseIf{$\neg$\istd{x} $\land$ \isod{y} $\land$ $\neg$\istran{y} $\land$ \isod{lhs}}{
	\tcp{e.g. w*points - dot product with sample features}
        {$\mathcal{P}_a(x) = REP$}
	}\uElseIf{$\neg$\isseq{x} $\land$ $\neg$\isseq{y} $\land$ $\neg$\isseq{lhs} $\land$
		(\istd{x} $\lor$ \istd{y} $\lor$ \istd{lhs}) }{
	\tcp{If any array is 2D, all arrays should be 2D}
        {$\mathcal{P}_a(lhs) = \mathcal{P}_a(x) = \mathcal{P}_a(y) = 2D\_BC$}
	}\uElse{}{
	\tcp{all replicated if no rule applied}
        {$\mathcal{P}_a(lhs) = \mathcal{P}_a(x) = \mathcal{P}_a(y) = REP$}\;
	}
\KwRet\;}
\caption{HPAT distribution analysis for matrix multiply.}
\label{fig:gemminfer}
\end{algorithm}

\begin{algorithm}
\DontPrintSemicolon
\SetKwFunction{procpp}{applyParforRules}
\SetKwProg{myproc}{Procedure}{}{}
\myproc{\procpp{parfor}}{
distribution $=$ 1D\;
myArrays $=$ $\emptyset$\;
arrayAccesses $=$ extractArrayAccesses(parfor.body)\;
parforIndexVar $=$ parfor.LoopNests[end].index\_var\;
\For{each arrayAccess $\in$ arrayAccesses} {
	\If{parforIndexVar $=$ arrayAccess.index\_expr[end]}{
		myArrays $=$ myArrays $\cup$ arrayAccess.array\;
                {$distribution = distribution \wedge \mathcal{P}_a(arrayAccess.array)$}
	}\If{parforIndexVar $\in$ arrayAccess.index\_expr[1:end-1]}{
		distribution $=$ REP\;
	}\If{isDependent(arrayAccess.index\_expr[1:end], parforIndexVar)}{
		distribution $=$ REP\;
	}
}
{$\mathcal{P}_p(parfor) = distribution$}\;
\For{each array $\in$ myArrays} {
{$\mathcal{P}_a(array) = distribution$}
}
\KwRet\;}
\caption{HPAT inference rules for parfors.}
\label{fig:parforinference}
\end{algorithm}

\noindent \emph{Transfer Function: Parfor - }
As described in Section~\ref{sec:distributed}, \emph{parfor} nodes represent data-parallel computation
and require special handling during distribution analysis.

\centerline{$f_{parfor x}: applyParforRules(x)$}

Figure~\ref{fig:parforinference} illustrates the transfer function for \emph{parfors}.
For clarity, this figure is simplified and only shows how the common case of \emph{1D\_B} arrays are handled.
As with array distribution, we start by assuming that the \emph{parfor} is \emph{1D\_B} until proven otherwise.
First, we analyze the body of the \emph{parfor} and extract all the array access/indexing operations.
We then iterate across all the array accesses.
Since HPAT parallelizes \emph{1D\_B} arrays and parfors across the last dimension,
the index variable of the last loop of the parfor is used for testing.
First, we check if the index expression for the last dimension of the array access (i.e., index\_exprN) is 
identical to the \emph{parfor's} index variable allocated to the last dimension of the loop nest.
If so and the array being accessed is \emph{REP} then the \emph{parfor} itself becomes \emph{REP} (the {\em meet} of two distributions is chosen).
Second, we check whether the last dimension's \emph{parfor} loop index variable is used
 directly or indirectly (e.g. temp = parfor.LoopsNests[end].index\_var; index\_expr1 = temp + 1) in any of the array access index expressions for the first N-1 dimensions.
If so, then the parfor must be \emph{REP}.
These tests are conservative but do not typically prevent parallelization of common analytics codes.
For accesses to \emph{2D\_BC} arrays, the above algorithm has a straight-forward extension that
considers not one but the last two parfor index variables and array index expressions.

\begin{algorithm}
\TitleOfAlgo{SGD}
\DontPrintSemicolon
samples $=$ read\_input()\;
w $=$ initial\_w()\;
\For{each iteration} {
	w $=$ gradient\_update(samples, w)\;
}
return w\;
\caption{Typical gradient descent method.}
\label{fig:sgd}
\end{algorithm}

\subsection{Effectiveness on Data Analytics Programs}\label{sec:sgd}

The main reason HPAT's heuristics are effective is that data analytics programs
typically produce a summary of large data sets.
In the case of machine learning algorithms, this summary is the weights of the trained model.
More specifically, many large-scale
machine learning algorithms
optimization methods such as
stochastic gradient descent (SGD)~\cite{bottou2010large}.
Hence, their structure can be represented as in Figure~\ref{fig:sgd}.
Parameter set \emph{w} is updated iteratively using gradient updates
in order to minimize a form of cost function on samples. HPAT's analysis can infer that \emph{samples}
is distributed since it will be accessed in a data parallel manner. It will also infer
that \emph{w} is replicated since it is updated using a form of reduction.
For example, variable \emph{w} in Figure~\ref{fig:hpatlr} is updated
by a matrix/vector multiplication that implies a reduction.

\subsection{Domain-specific Optimizations}\label{sec:domainopt}
Before distributed code generation, HPAT applies two novel domain-specific optimization heuristics.
HPAT performs these since typical compiler cost models cannot reliably conclude that the complex transformations described below
are beneficial; for example, we found ICC and GCC unable to perform these optimizations on any of our benchmarks.
Hence, HPAT's use of domain knowledge is critical.
For the first heuristic, we observe that matrix multiplications (GEMM calls) with at least one ``tall-and-skinny''
input are common in data analytics algorithms (e.g. $w*points$ in Figure~\ref{fig:lrcode}).
Ideally, these GEMM calls should be replaced with equivalent loop-nests and fused with other operations
to achieve a single pass through data points and intermediate results:

\begin{heuristic}
	GEMM operations with one or more 1D\_B inputs are replaced with loop-nests
	and fusion is called on the basic block.
\end{heuristic}

To enable fusion, HPAT arranges the loop-nests
so that the long dimension is outer-most.
This optimization causes all mathematical operations of the logistic regression algorithm
in Figure~\ref{fig:lrcode} to be fused.
Hence, each data point
is loaded just once, which improves performance by maximizing locality.
Furthermore, memory consumption is improved since saving intermediate data
into memory is avoided.

The second heuristic is based on the observation that loops over data sets and
their intermediate results can occur inside other loops.
For example, the {\em centroids} calculation in Figure~\ref{fig:hpatkmeans} is written
using nested comprehensions that include passes over {\em points} and {\em labels} inside.
This prevents maximum loop fusion and causes multiple passes over those data sets.
Furthermore, extra communication is then generated for
each iteration of the outer loop-nest instead
of a singe communication call for all data exchanges.
Thus, we rearrange loop-nests to avoid this problem:

\begin{heuristic}
	Parfors with REP distribution that have 1D\_B parfors inside are interchanged
	and fusion is called on the basic block.
\end{heuristic}

HPAT performs loop fission on the REP parfor before interchange
since the body may have more statements.
This optimization maximizes fusion in the kmeans example of Figure~\ref{fig:hpatkmeans}
and ensures a single pass over the data set.

\begin{figure}
\begin{lstlisting}[language=Julia,escapeinside={(*}{*)},mathescape]
using HPAT

@acc hpat function kmeans(numCenter, iters, file)
  points = DataSource(Matrix{Float64},HDF5,"points",file)
  D,N = size(points)
  centroids = rand(D, numCenter)
  for l in 1:iters
    dist = [Float64[sqrt(sum((points[:,i]-centroids[:,j]).^2)) for j in 1:numCenter] for i in 1:N]
    labels = Int[indmin(dist[i]) for i in 1:N]
    centroids = Float64[ sum(points[j,labels.==i])/      sum(labels.==i) for j in 1:D, i in 1:numCenter]
    end
    return centroids
end

centroids = kmeans(5,100,"mydata.hdf5")
\end{lstlisting}
\caption{HPAT K-Means example.}
\label{fig:hpatkmeans}
\end{figure}


\subsection{Automatic Parallel I/O}\label{sec:datasource}

\begin{figure}
\centering
\includegraphics[width=\columnwidth]{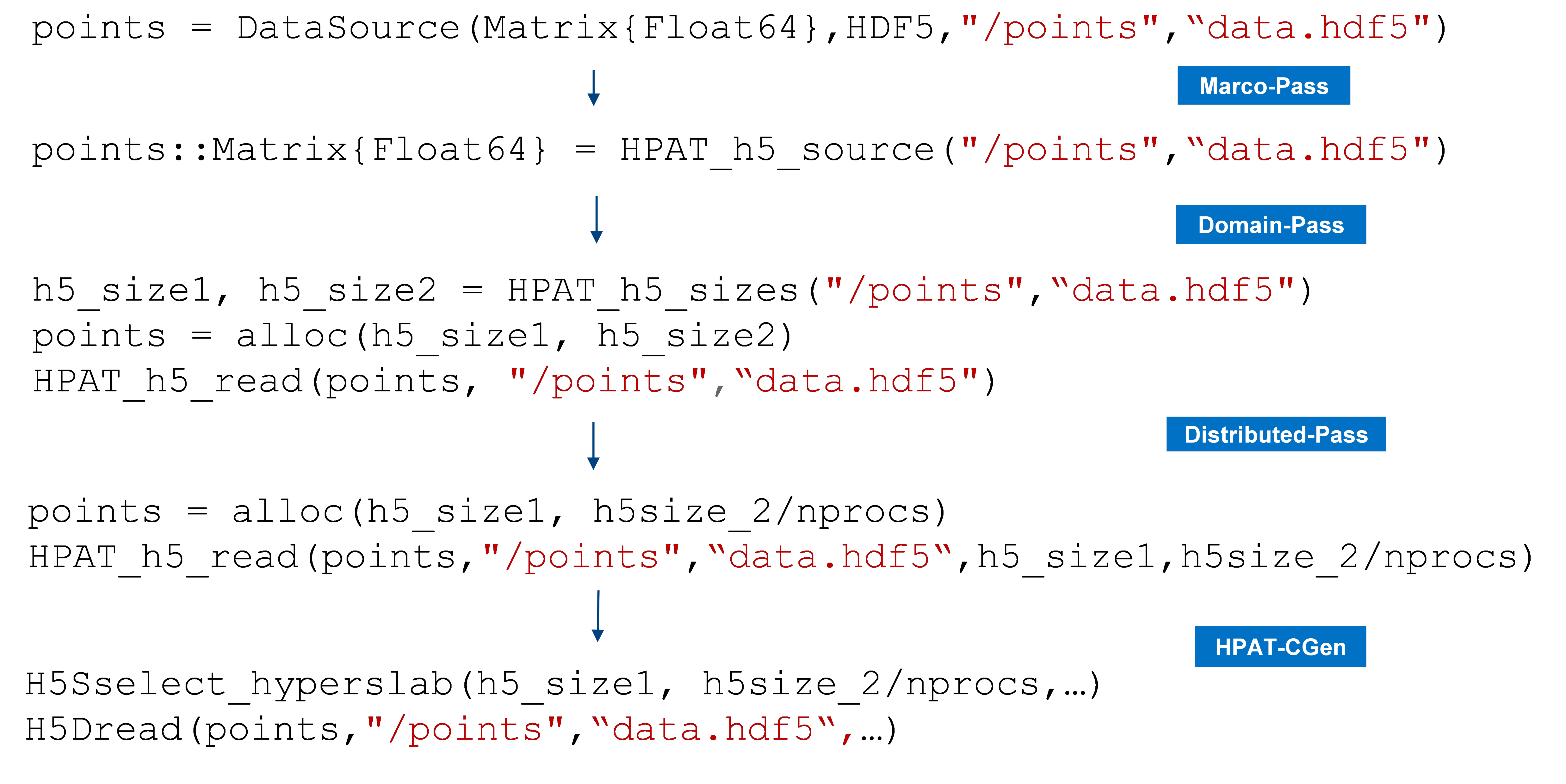}
\caption{HPAT's DataSource compilation pipeline.}
\label{fig:datasource}
\end{figure}

Figure~\ref{fig:datasource} demonstrates how HPAT's compilation pipeline translates
 \emph{DataSource} syntax to parallel I/O code (\emph{DataSink} is similar).
Macro-Pass desugars the syntax into a HPAT placeholder special function call (e.g. \emph{HPAT\_h5\_read})
and type-annotates the output array so that Julia's
type inference can work.
Domain-Pass generates function calls to get the size of the array and
allocations so that Domain-IR and Parallel-IR
can optimize the program effectively.
Allocations and size variables are critical information that enable fusion and elimination of intermediate arrays.

Distributed-Pass enables parallel I/O by distributing the input array among nodes
and adding the start and end indices for each dimension.
HPAT Code Generation's function call replacement mechanism generates
the backend HDF5 code (currently MPI/C++) for placeholders such as \emph{HPAT\_h5\_read}.
HPAT also supports text files using MPI I/O because many big data files are stored as text.

\subsection{Distributed-Memory Translation}\label{sec:distributedpass}

Distributed-Pass transforms the function for distributed-memory execution and generates communication calls
after auto-parallelization provides distributions.
%
The pass also inserts calls into the IR to query the number of processors and
to get the node number. 
Allocations of distributed arrays are divided among nodes by inserting code
to calculate size based on the number of processors (e.g. $mysize=total/num\_pes$).
Distributed (\emph{1D\_B}) parfors are translated by dividing the iterations among processors using node number and number of processors
and updating array indices accordingly.
For example, $A[i]$ is replaced with $A[i-mystart]$ where $mystart$ contains
the starting index of the parfor on the current processor.
Furthermore, for parfors with reductions, communication calls
for distributed-memory reductions are generated.

Furthermore, matrix/vector and matrix/matrix multiplication (GEMM calls) need special handling.
For instance, $w*points$ in Figure~\ref{fig:hpatlr} does not require communication since replication of \emph{w} makes it data-parallel,
 while $(\cdots.*labels)*points'$
requires an \emph{allreduce} operation since both arguments are distributed.
The Distributed-Pass makes this distinction by using the parallelization information provided by previous analyses.
In the first case, the first input is replicated while the second input and the output are distributed.
In the second case, both inputs of GEMM are distributed but the output is replicated.

\subsection{Backend Code Generation}\label{sec:backend}
Our approach facilitates using various backends, since Distributed-Pass returns a high-level
parallel IR that enables flexible code generation.
Our current prototype supports \mpic and
\mpiompc, which we found to be suitable for many analytics workloads.
However, some cases might require using a backend with an adaptive runtime system,
due to scheduling and load balancing requirements. Hence, one might use
\charm~\cite{sc14charm} or HPX~\cite{Kaiser:2014:HPX}
as the backend for HPAT.
Evaluation of these backends for HPAT is left for future work.

On the other hand, Spark uses a runtime system with DAG scheduling,
which is required for implementation of complex operations
(such as shuffling) on top of Spark's map/reduce core.
However, these operations do not necessarily need runtime scheduling.
In general, Spark surveys show that only a small fraction of users run
irregular workloads such as graph workloads on Spark~\cite{sparksurvey2015,sparksurvey2016}.
Nevertheless, our approach enables the use of low-overhead HPC runtime systems,
avoiding Spark's overheads.

\subsection{Automatic Utilization of Distributed-Memory Libraries}\label{sec:library}
Libraries are essential components of productive analytics platforms.
For example, Spark provides MLlib~\cite{mllibUrl}, which is implemented
using its RDD data format.
Since HPAT is compiler based, it can take advantage
of distributed-memory libraries without
requiring changes to their data structures and interfaces.
On the other hand, only libraries that implement
interoperation with Spark's RDDs can be used with Spark;
this is a fundamental limitation for library-based frameworks.

\begin{figure}
\begin{lstlisting}[language=Julia,escapeinside={(*}{*)},mathescape,numbers=none]
using HPAT

@acc hpat function calcNaiveBayes(num_classes, file_name)
    points = DataSource(Matrix{Float64},HDF5,"/points",file_name)
    labels = DataSource(Vector{Float64},HDF5,"/labels", file_name)
    coeffs = HPAT.NaiveBayes(points, labels, num_classes)
    return coeffs
end
\end{lstlisting}
\caption{HPAT library call example.}
\label{fig:hpatklib}
\end{figure}

Figure~\ref{fig:hpatklib} shows example code that
calls a library. This function call goes through
the compilation pipeline as a special known function call, i.e., it has
entries in HPAT analysis tables.
Most importantly, the HPAT parallelization algorithm knows
that the input arrays can be {\em 1D\_B}, while
the output array is {\em REP}.
If parallelization is successful, the Distributed-Pass
adds two new arguments to the function call; the first index
and the last index of the input array on each node.
HPAT's CGen extension uses a \mpic code
routines for code generation of the call.
Currently, HPAT supports ScaLAPACK and
\INTELR~Data Analytics Acceleration Library (\INTELR~DAAL)~\cite{daalUrl}
as an alternative to MLlib for machine learning algorithms.

Performance evaluation of analytics programs that use libraries
(e.g. MLlib or DAAL) is beyond the scope
of this paper. Comparing libraries is non-trivial in general;
data analytics functions can be implemented using multiple algorithms with
different computational complexities.
Our objective is to generate efficient code for scripting codes
which are dominantly used in this area.

% ...


\subsection{2D Parallelization}\label{sec:library}

\begin{figure}
\begin{lstlisting}[language=Julia,escapeinside={(*}{*)},mathescape]
using HPAT

@acc hpat function matrix_multiply(file1,file2,file3)
  @partitioned(M,2D);
  M = DataSource(Matrix{Float64},HDF5,"/M", file1)
  x = DataSource(Matrix{Float64},HDF5,"/x", file2)
  y = M*x
  y += 0.1*randn(size(y))
  DataSink(y,HDF5,"/y", file3)
end

\end{lstlisting}
\caption{HPAT matrix multiply example.}
\label{fig:hpatpmm}
\end{figure}

HPAT requires a simple annotation to assist the automatic parallelization algorithm
for the less common cases that require 2D parallelization.
Consider the example program in Figure~\ref{fig:hpatpmm} (a real-world case requested by a user).
The program reads two (multi-terabyte) matrices, multiplies them,
adds some random value to the result, and writes it back to storage.
The user specifies that matrix \emph{M} requires 2D partitioning. HPAT infers that
\emph{x} and \emph{y} matrices also need 2D partitioning as well. Furthermore, the related intermediate variables
in the AST (such as the random matrix created) and the parfors are also 2D partitioned.
In the backend, HPAT generates MPI/C++ code which calls parallel HDF5 for I/O and ScaLAPACK (PBLAS component)
for matrix multiplication.

Manually developing efficient parallel code for this program is challenging.
ScaLAPACK requires block-cyclic partitioning of input and output data but
HDF5 provides a block-based read/write interface for parallel I/O.
Hence, the generated code has to sets 96 indices, since there are three I/O operations;
each operation requires four hyperslab selections including corner cases,
 and each hyperslab selection requires start, stride, count and block size indices for two dimensions.
In addition, ScaLAPACK's legacy interface is Fortran-based and has its own intricacies.
As a result, the generated MPI/C++ code is 525 lines.
Therefore, manually developing this code is highly error-prone
for domain experts and HPAT's automation is significant.

This use case demonstrates a fundamental advantage of our compiler approach.
Library-based frameworks are based on fixed distributed data structures with specific partitioning and layout formats.
For example, Spark's RDDs are 1D partitioned and the whole framework
(e.g block manager and scheduler) is based on this 1D partitioning.
Supporting different partitionings
and layouts is easy for a compiler, but is difficult for a library.




\section{Checkpointing}\label{sec:checkpointing}
Resiliency is essential for long-running iterative machine learning algorithms.
The challenge is to provide low-overhead fault tolerance support without significant
programmer involvement.
Previous work on automatic checkpointing cannot achieve minimal checkpoints
since, for example, those systems do not have the high-level knowledge that
models are replicated and one copy is enough~\cite{Schulz:2004:IES,pbk:95:came,Wang:2008:AAC}.
Spark's approach is to use lineage to restart shortest possible tasks but
this is made possible only by a system design with high overheads.
HPAT provides automated application-level checkpointing (ALC)
based on domain characteristics.

For iterative machine learning applications, only the learning parameters
and the loop index need to be checkpointed since the data points are read-only.
Moreover, these learning parameters are replicated
on all processors so only one copy needs to be checkpointed.
We use these domain characteristic in designing HPAT's checkpointing capability.
Hence, HPAT checkpointing assumes a typical analytics function containing a single outer-loop and having the form: initialization, iteration loop, results.
In the initialization phase, input is loaded and variables initialized, which establish the invariants for entry into the loop.
The body of the outer-loop can be arbitrarily complex including containing nested loops.
In the results phase, the outputs of the loop are used to compute the final result of the function.
This approach is readily extensible to support multiple outer-loops as well.

If enabled, HPAT adds a checkpointing pass to the compilation pipeline after Domain-Pass.
The checkpointing pass first locates the outer-loop and analyzes it to determine
which variables are live at entry to the loop (including the loop index) and are written in the loop.
This set of iteration-dependent variables are saved as part of a checkpoint.
The pass creates a new checkpoint function tailored to this set of variables and inserts a call to that function as the first statement of the loop.
In this function, MPI rank zero compares the time since the last checkpoint was taken with the next checkpoint time as calculated using Young's formula~\cite{Young}.
If it is time to take a checkpoint, rank zero calls the HPAT runtime to start a checkpointing session, write each variable to the checkpoint, and then end the session.
The HPAT runtime records the time to take the checkpoint and uses this information to improve the estimated checkpoint time that is input to Young's formula.
At the exit of the outer-loop, the pass also inserts a call to the HPAT runtime to indicate that the checkpointed region has completed and that any saved checkpoints can be deleted.

To restart a computation, the programmer calls the HPAT restart routine and passes the function to be restarted and the original arguments.
The HPAT compiler creates a restart version of the function that is identical to the original but with the addition of checkpoint restore code before the entry to the loop.
This checkpoint restore code finds the saved checkpoint file and loads the iteration-dependent variables from the checkpoint.
In this way, the initialization code is performed again (restoring read-only variables),
 and the loop fast-forwards to the last successfully checkpointed iteration.
An important consideration is that the iterations should be deterministic since
some might be re-executed during restart.

Consider the logistic regression example in Figure~\ref{fig:hpatlr}; we store only the loop index \emph{i}
and \emph{w} in the checkpoint whereas the full set of live data would include \emph{points} and \emph{labels} and would result in checkpoints orders of magnitude larger.
As far as we are aware, this checkpointing approach that exploits domain knowledge
by for example re-executing the initialization phase is novel.
A key insight is that HPAT can perform the analysis for this minimal checkpointing,
while a library approach like Spark is unable to do so.


\section{ParallelAccelerator Infrastructure}\label{sec:parallelaccelerator}
HPAT relies on the ParallelAccelerator package~\cite{paGithub} for discovering potential parallelism in dense array operations of Julia.
ParallelAccelerator consists of three main compiler passes, Domain-IR, Parallel-IR, and CGen.
Domain-IR looks for operations and other constructs in Julia's IR that have different kinds of parallel semantics and then replaces those operations with
equivalent Domain-IR nodes that encode those semantics.
Some of the most common parallelism patterns in Domain-IR are
\emph{map}, \emph{reduce}, \emph{Cartesian map}, and \emph{stencil}.
For example, Domain-IR would identify unary vector operations (such as -, !, log, exp, sin, and cos) and binary, element-wise vector-vector or
vector-scalar operations (such as +, -, *, /, ==, !=, \textless, and \textgreater) as having \emph{map} semantics.
Likewise, Julia's \emph{sum()} and \emph{prod()} functions would be identified by Domain-IR as having \emph{reduce} semantics.
Domain-IR identifies comprehensions within the Julia code as having \emph{Cartesian map} semantics.

The Parallel-IR pass lowers Domain-IR nodes to a common representation called \emph{parfor}.
Once in this representation, Parallel-IR performs parfor fusion between parfors coming from potentially dissimilar Domain-IR nodes.
This fusion process reduces loop overhead and eliminates many intermediate arrays, helping the program to have better locality.
There are three main components of the \emph{parfor} representation: loop nests, reductions, and body.
Every \emph{parfor} has a loop nest that represents a set of tightly nested for loops.
It is typical for the number of such loops to match the number of dimensions of the array on which the \emph{parfor} operates.
The \emph{parfor} reductions component is only present when the \emph{parfor}
involves a reduction and encodes the variable holding the reduction value along
with its initial value and the function used to combine reduction elements.
Finally, the body of the \emph{parfor} contains code to compute the result of
 the \emph{parfor} for a single point in the iteration space.
After the Parallel-IR pass is finished, CGen converts the IR to OpenMP-annotated C code in the backend.


\section{Evaluation}\label{sec:evaluation}

\begin{figure*}[ht!]
\begin{subfigure}{.45\linewidth}
    \centering
    \def\svgwidth{\columnwidth}
    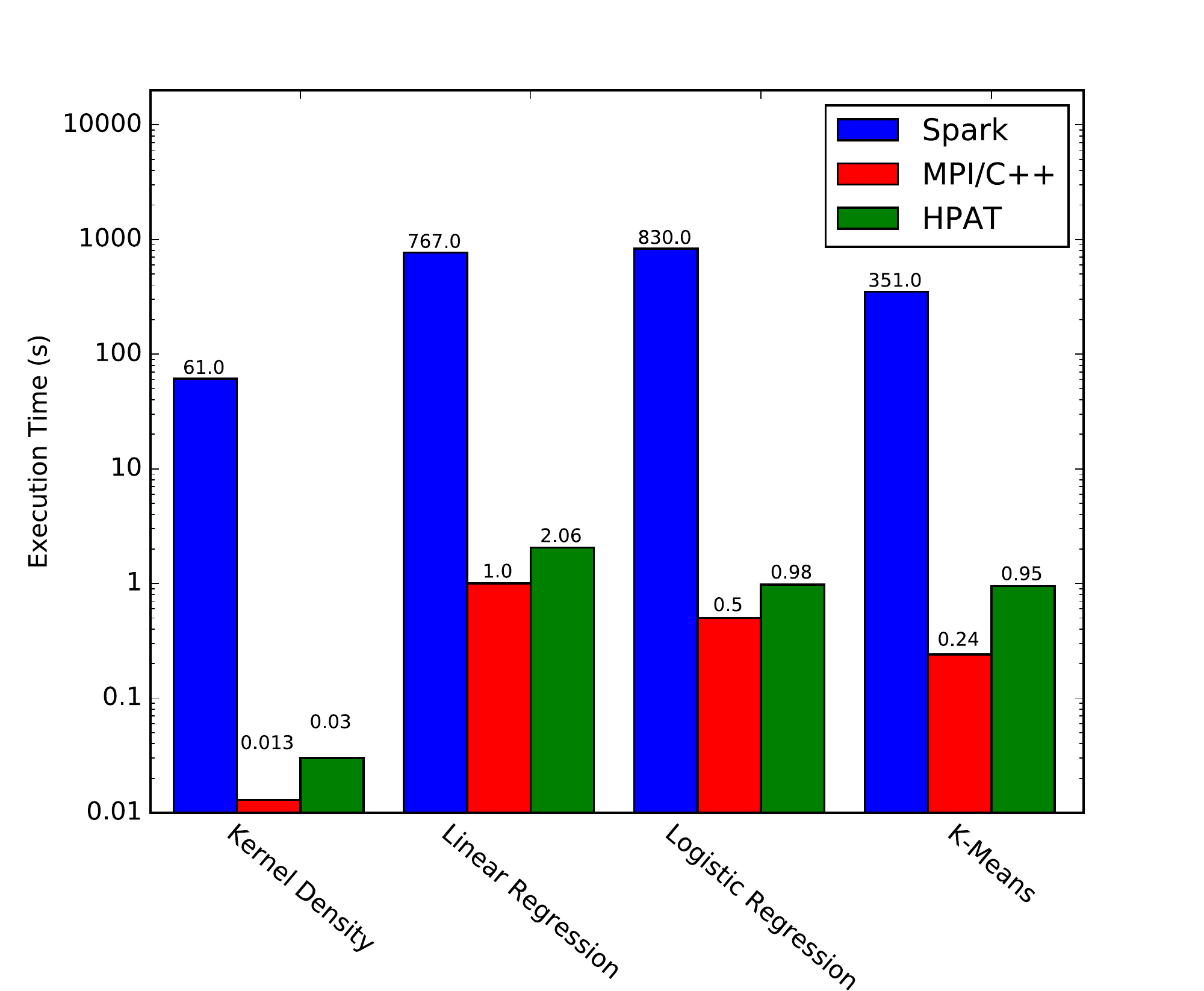
    \caption{Cori, 64 nodes.}
    \label{fig:cori-perf}
\end{subfigure}
\begin{subfigure}{.45\linewidth}
    \centering
    \def\svgwidth{\columnwidth}
    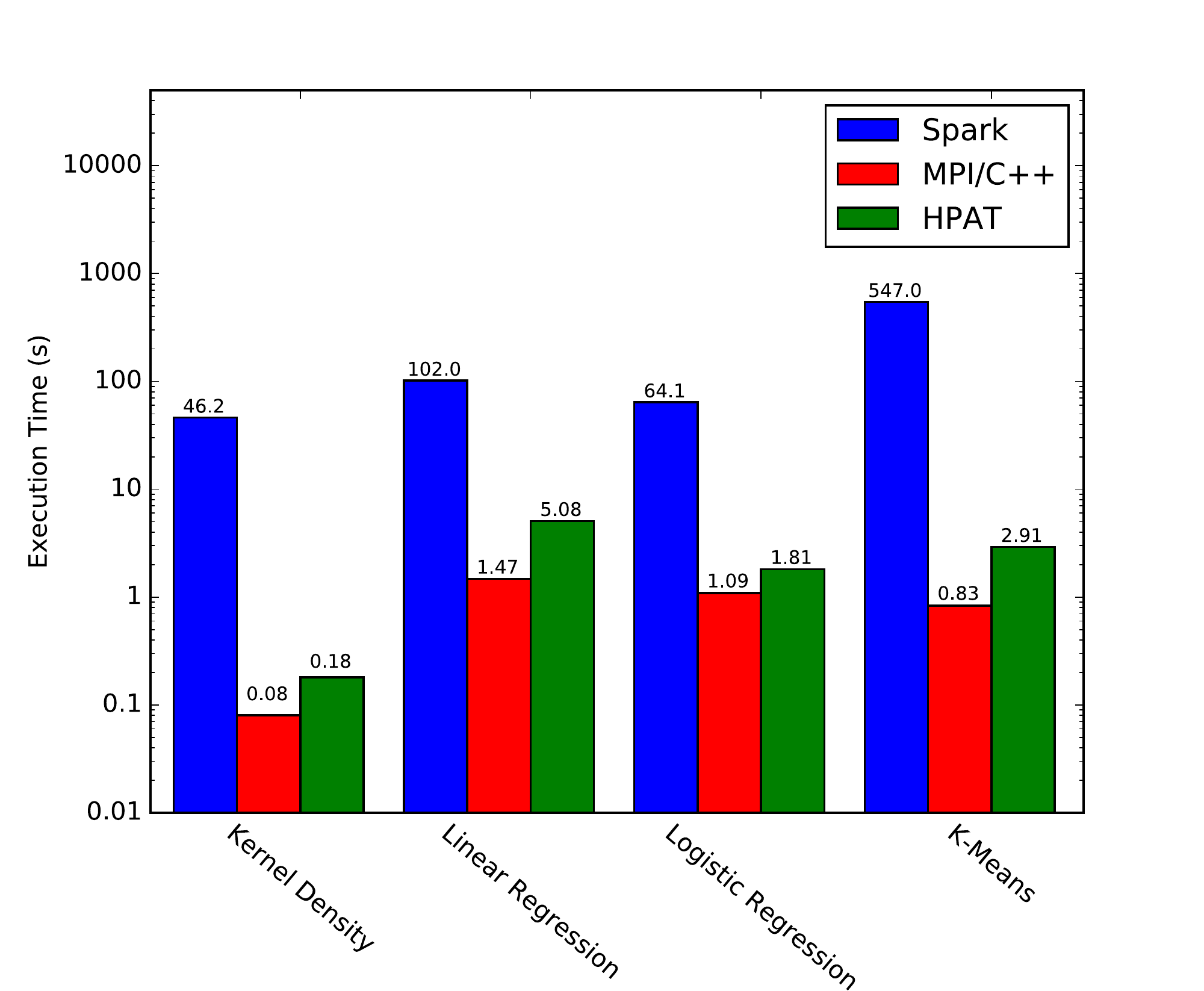
    \caption{Amazon AWS, 4 instances (c4.8xlarge).}
    \label{fig:aws-perf}
\end{subfigure}
\caption{Performance comparison of Spark, manual MPI/C++, and HPAT. Note the logarithmic scales.}
\label{fig:perf}
\end{figure*}

We compare the performance of Spark, HPAT and handwritten MPI/C++ programs on the Cori supercomputer
at NERSC~\cite{coriNERSC} and Amazon AWS cloud.
Cori is a Cray XC40 supercomputer that
includes advanced features for data-intensive applications such as
large memory capacity and high I/O bandwidth.
Each node (Phase I) has two sockets, each of which is
a 16-core \INTELR \XEONR E5-2698 v3 \@ 2.3GHz ({\em Haswell} architecture).
The memory capacity of each node is 128GB.
On AWS, we use the {\em compute-optimized} C4 instances (c4.8xlarge with 36 vCPUs),
 which feature \INTELR \XEONR E5-2666 v3 (Haswell) processors.
Each instance has 60GB of memory. We use Placement Groups which provide
low latency, full bisection, 10Gbps bandwidth between instances.
We use \INTELR \cpp Compiler (ICC) 17.0 for backend \cpp compilation.
%
%
%
%
The default Spark 2.0.0 installation on Cori is used which is tuned and supported.

We use the benchmarks listed in Table~\ref{table:benchmarks} for evaluation.
We believe they are representative of many workloads in data analytics;
related studies typically use the same or similar benchmarks~\cite{brown2016have,
clusterApache15,OusterhoutAnalysis15,ReyesOrtizMPIspark2015}.
Benchmark sizes are chosen so that they fit in the memory, even
with excessive memory usage of Spark, to make sure Spark's performance is not degraded
by accessing disks repeatedly.
HPAT is capable of generating MPI/OpenMP but currently, it turns
OpenMP off and uses MPI-only configuration since OpenMP code generation is
not tuned yet. 
We use one MPI rank per core (without hyperthreading).
The Spark versions of the benchmarks are based on the available examples in Spark's open-source distribution.
We developed the MPI/C++ programs by
 simply dividing the problem domain across ranks equally
and ensuring maximum locality by fusing the loops manually.
Parallel I/O times are excluded from all results.
\mpic codes are about 6$\times$ longer in lines of code compared to HPAT codes.

\begin{table}
\footnotesize
\caption{Benchmark sizes and parameters.}
\begin{minipage}{\linewidth}
\begin{tabular}{l l l c r}
\toprule
\textbf{Benchmark} & \textbf{Input size/iterations on Cori}   & \textbf{AWS}    \\ \midrule
Kernel Density         &  2B/(NA) & 256M/(NA)  \\
Linear Regression             & 2B/20 (10 features, 4 models)  & 256M/20  \\
Logistic Regression   & 2B/20 (10 features) & 256M/20 \\
K-Means   & 320M/20 (10 features, 5 centroids) & 64M/20 \\
\end{tabular}
\end{minipage}
\label{table:benchmarks}
\end{table}

Figure~\ref{fig:perf} compares the performance of Spark, manual MPI/C++, and HPAT on 64 nodes (2048 cores) of Cori
and four nodes of Amazon AWS.
HPAT is 369$\times$-2033$\times$ faster than Spark for the benchmarks on Cori
and is only 2$\times$-4$\times$ slower than manual MPI/C++.
In addition, HPAT is 20$\times$-256$\times$ faster than Spark on Amazon AWS.
The lower performance of Spark on Cori is expected since the master node is a bottleneck and prevents scaling.
\emph{Kernel Density} demonstrates the largest performance gap among the benchmarks; HPAT is 2033$\times$ faster than Spark on Cori and 256$\times$ on AWS.
The reason is that computation per element is small
and the Spark overheads such as serialization/deserialization and master-executor coordination are amplified.

Automatic parallelization by HPAT matches the manual parallelization for all of the benchmarks perfectly,
which demonstrates the robustness of our auto-parallelization algorithm.
Furthermore, loop structures are identical to the manual versions which demonstrates
the success of HPAT's fusion transformations.
The performance gap between HPAT and \mpic codes can be attributed to the
the generated code being longer and containing extra intermediate variables;
this makes code generation phase of the backend \cpp compiler more challenging.
The optimization reports of ICC support this hypothesis.
We hope to match the performance of manual codes in future work.

We use the Python interface of Spark in this paper as baseline since scripting languages are
preferred by domain experts. One may argue that using Scala,
which is statically compiled, might be significantly faster in Spark.
We tested this hypothesis
 and found that Scala is moderately faster for some benchmarks
(e.g. {\em K-Means}) but HPAT is still several times faster than Spark.
For other benchmarks (e.g. {\em Logistic Regression}) Python is actually faster
since the computation is inside Numpy operations, which have optimized native backends (Anaconda distribution).
Furthermore, the flexibility of Numpy allows batched processing while Scala's {\em breeze} linear algebra
library does not have this capability.

\begin{figure}[ht!]
    \centering
    \def\svgwidth{.9\columnwidth}
    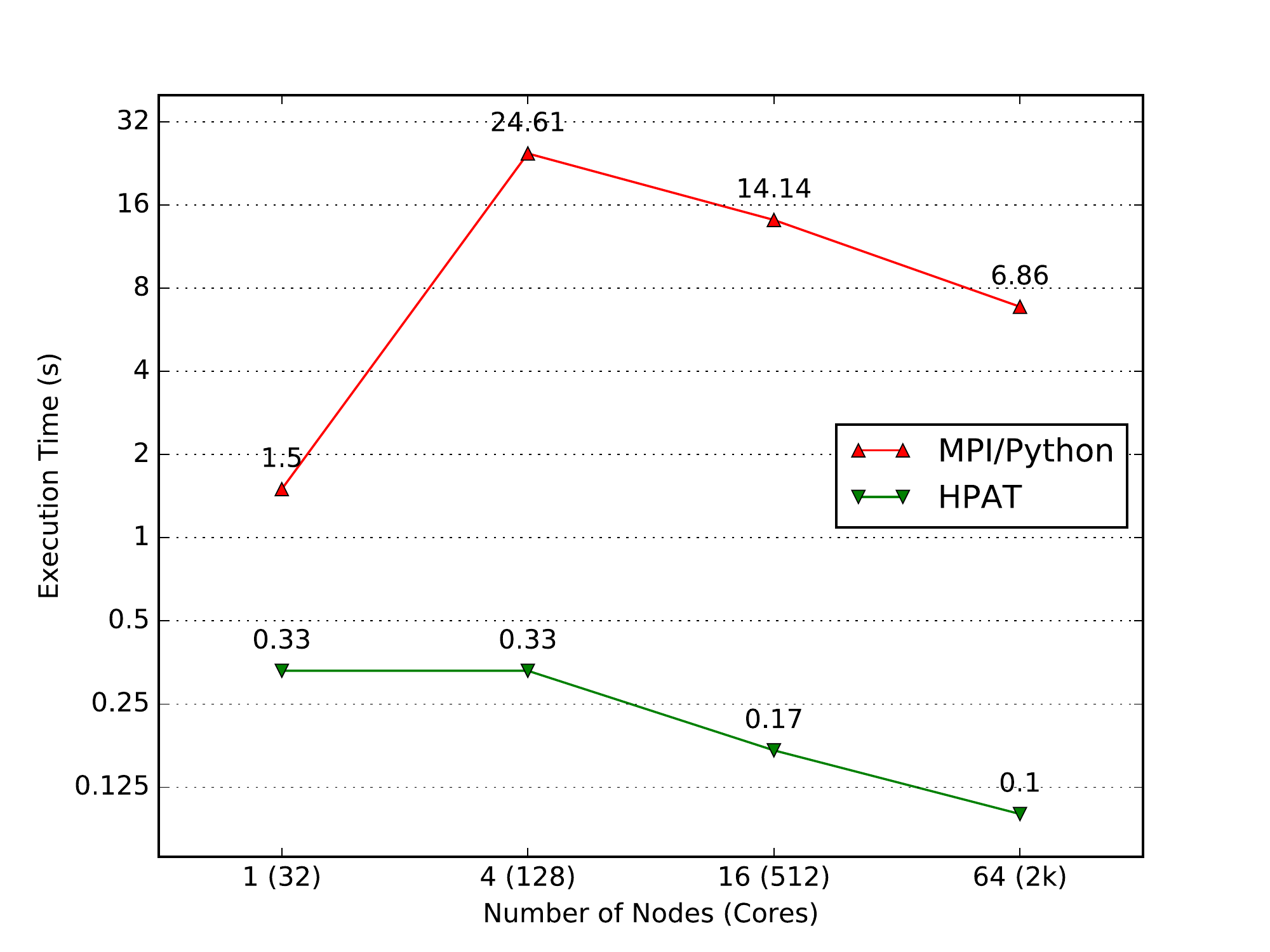
    \caption{Strong scaling of ADMM LASSO. Note the logarithmic scales.}
    \label{fig:admm}
\end{figure}

We use an {\em ADMM LASSO} solver~\cite{wahlberg2012admm} to evaluate the effectiveness of our auto-parallelization
method on a complex algorithm. The code is originally written in Python and parallelized using {\em mpi4py} by
a domain expert. However, the manual parallelization method sacrifices accuracy for easier parallelization.
On the other hand, HPAT is able to parallelize the code effectively and accurately.
Figure~\ref{fig:admm} demonstrates the scaling on up to 64 nodes.
The slowdown of the MPI/Python code running in parallel is partially due to accuracy loss which forces the
algorithm to run up to specified maximum number of iterations.
The success of HPAT's auto-parallelization on this algorithm provides confidence about the effectiveness of our approach,
since we do not expect analytics algorithms to be significantly more complex than this case in practice.


\paragraph{Compiler feedback and control:}
Previous compiler approaches are hard to understand and control by users since
they use complex heuristics and cost models.
HPAT's automatic parallelization algorithm inherently facilitates user control since it is deterministic
and can be controlled easily.
For example, HPAT provides the operations that caused each REP inference.
 The user is then able to change parallelization behavior by
explicitly annotating parallelization for arrays or
 providing more information for operations (e.g. parallelization for library calls
not previously known to HPAT).


\section{Related Work}\label{sec:related}

Previous studies demonstrate that current big data analytics frameworks are orders of magnitude slower than
hand-tuned MPI implementations~\cite{brown2016have,ReyesOrtizMPIspark2015,mcsherry2015scalability}.
To improve the performance of data analytics frameworks, some previous studies
have focused on more efficient inter-node communication~\cite{clusterApache15,zhang2015harp} but
they do not address the fundamental performance bottlenecks resulting
from the library approach.
HPAT solves this problem using a novel automatic parallelization approach.

Automatic parallelization is studied extensively in HPC, but
it is known that auto-parallelizing
compilers cannot match the performance of hand-written
parallel programs for many applications~\cite{Eigenmann94onthe,Wilson:1994:SIR,Blume:1996:PPP,Waheed99cap,Hiranandani:1994:DEN}.
For example, previous studies have tried to automate data alignment (TEMPLATE and ALIGN directives)
and distribution (DISTRIBUTE directive) steps in High Performance Fortran (HPF) with
limited success~\cite{Kennedy:2007:RFH,Chatterjee:1995:OEA,Kennedy:1998:ADL}.
More specifically, our distribution analysis algorithm can be compared
with the framework by Kennedy and Kremer~\cite{Kennedy:1998:ADL}.
This framework performs a search of possible alignment and distribution combinations for
 loop-nests; a performance model helps predict the combination with the best performance.
However, this framework cannot find the best combination reliably due to the inaccuracies of the performance model.
Conversely, the HPAT algorithm leverages domain knowledge
and assigns distributions to arrays and computations based on semantics of different high-level operations.
To the best of our knowledge, this is the only algorithm to use
a data flow formulation and is therefore novel.
In general, previous work targeting scientific computing applications relies on
complex analysis, performance models and approximations
that are known to have challenges in a practical setting~\cite{Kennedy:2007:RFH,Waheed99cap}.
In contrast, our algorithm does not rely on models and approximations,
and is therefore accurate (matches manual parallelization).

Distributed Multiloop Language (DMLL) provides compiler transformations on
map-reduce programs on distributed heterogeneous architectures~\cite{brown2016have}.
However, HPAT starts from higher level scripting code,
which might have operations like matrix-multiply that do not have a simple map-reduce translation.
More importantly, DMLL relies on user annotations and a simple partitioning analysis pass for
data partitioning but HPAT is fully automatic.
We believe that our iterative partitioning analysis algorithm produces
more efficient code since it does not parallelize all potentially parallel operations like DMLL's approach.
This can affect communication cost on distributed-memory operations significantly since DMLL broadcasts
local data structures.
This could be the reason for DMLL's much smaller speedups over Spark on CPU clusters.

Distributed Halide is a domain-specific compiler for image processing that translates high-level stencil pipelines into
parallel code for distributed-memory architectures~\cite{denniston2016distributed}.
Unlike HPAT, 2D partitioning and near neighbor communication are the norm and not 1D partitioning and reductions.
Moreover, Halide requires user ``schedule'' for optimization and distributed execution
while HPAT is fully automatic.

Several previous efforts such as Delite~\cite{Sujeeth:2014:DCA},
Copperhead~\cite{Catanzaro:2011:CCE}, and \INTELR Array Building Blocks~\cite{Newburn:2011:IAB}
provide embedded domain-specific languages (DSLs) that are compiled for parallel hardware.
HPAT's design is similar in many aspects, but HPAT targets data analytics for
distributed-memory architectures (with more accurate auto-parallelization).
Furthermore, HPAT uses the abstractions of the host language and avoids introducing new
abstractions to the programmer as much as possible.

Systems that automate application-level checkpointing to some degree have been proposed before~\cite{Schulz:2004:IES,pbk:95:came,Wang:2008:AAC}.
For example, in the method by Plank et al.~\cite{pbk:95:came}, the programmer adds checkpoint
locations and the system excludes dead variables and read-only data from checkpoints.
In contrast, HPAT's scheme is domain-specific and, for example, uses the knowledge that
 the learning parameters in HPAT are replicated;
 checkpointing them does not require an MPI consistency protocol.
HPAT also completely automates the checkpointing process
whereas other systems require programmer effort to some degree.

\section{Conclusion and Future Work}\label{sec:conclusion}
Library-based big data analytics frameworks such as Spark provide
programmer productivity but they are much slower
than hand-tuned MPI/C++ codes due to immense runtime overheads.
We introduced an alternative approach based on a novel auto-parallelization algorithm,
 which is implemented in High Performance Analytics Toolkit (HPAT).
HPAT provides the best of both worlds: productivity of scripting abstractions
and performance of efficient MPI/C++ codes.
Our evaluation demonstrated that HPAT is 369$\times$-2033$\times$ faster than Spark.
We plan to expand HPAT to provide more data analytics features and use cases.
For example, providing support for
 sparse computations, data frames (heterogeneous tables),
 out-of-core execution is under investigation.
Most of these features need research on multiple layers; from scripting abstractions to compilation techniques
and code generation.
We are also building a prototype HPAT system for Python.


\bibliographystyle{ACM-Reference-Format}
\bibliography{hpat}


\begin{thebibliography}{00}


\ifx \showCODEN    \undefined \def \showCODEN     #1{\unskip}     \fi
\ifx \showDOI      \undefined \def \showDOI       #1{{\tt DOI:}\penalty0{#1}\ }
  \fi
\ifx \showISBNx    \undefined \def \showISBNx     #1{\unskip}     \fi
\ifx \showISBNxiii \undefined \def \showISBNxiii  #1{\unskip}     \fi
\ifx \showISSN     \undefined \def \showISSN      #1{\unskip}     \fi
\ifx \showLCCN     \undefined \def \showLCCN      #1{\unskip}     \fi
\ifx \shownote     \undefined \def \shownote      #1{#1}          \fi
\ifx \showarticletitle \undefined \def \showarticletitle #1{#1}   \fi
\ifx \showURL      \undefined \def \showURL       #1{#1}          \fi
\providecommand\bibfield[2]{#2}
\providecommand\bibinfo[2]{#2}
\providecommand\natexlab[1]{#1}
\providecommand\showeprint[2][]{arXiv:#2}

\bibitem[\protect\citeauthoryear{??}{spa}{2015}]%
        {sparksurvey2015}
 \bibinfo{year}{2015}\natexlab{}.
\newblock \bibinfo{title}{{Apache Spark Survey 2015 Report}}.
\newblock
  \bibinfo{howpublished}{\url{http://go.databricks.com/2015-spark-survey/}}.
  (\bibinfo{year}{2015}).
\newblock


\bibitem[\protect\citeauthoryear{??}{spa}{2016}]%
        {sparksurvey2016}
 \bibinfo{year}{2016}\natexlab{}.
\newblock \bibinfo{title}{{Apache Spark Survey 2016 Report}}.
\newblock
  \bibinfo{howpublished}{\url{http://go.databricks.com/2016-spark-survey/}}.
  (\bibinfo{year}{2016}).
\newblock


\bibitem[\protect\citeauthoryear{??}{cor}{2016}]%
        {coriNERSC}
 \bibinfo{year}{2016}\natexlab{}.
\newblock \bibinfo{title}{{Cori Supercomputer at NERSC}}.
\newblock
  \bibinfo{howpublished}{\url{http://www.nersc.gov/users/computational-systems/cori/}}.
    (\bibinfo{year}{2016}).
\newblock


\bibitem[\protect\citeauthoryear{??}{daa}{2016}]%
        {daalUrl}
 \bibinfo{year}{2016}\natexlab{}.
\newblock \bibinfo{title}{{Intel Data Analytics Acceleration Library}}.
\newblock \bibinfo{howpublished}{\url{https://software.intel.com/en-us/daal}}.
   (\bibinfo{year}{2016}).
\newblock


\bibitem[\protect\citeauthoryear{??}{paG}{2016}]%
        {paGithub}
 \bibinfo{year}{2016}\natexlab{}.
\newblock \bibinfo{title}{{ParallelAccelerator Julia package}}.
\newblock
  \bibinfo{howpublished}{\url{https://github.com/IntelLabs/ParallelAccelerator.jl}}.
    (\bibinfo{year}{2016}).
\newblock


\bibitem[\protect\citeauthoryear{??}{mll}{2016}]%
        {mllibUrl}
 \bibinfo{year}{2016}\natexlab{}.
\newblock \bibinfo{title}{{Spark Machine Learning Library (MLlib) Guide}}.
\newblock
  \bibinfo{howpublished}{\url{http://spark.apache.org/docs/latest/mllib-guide.html}}.
    (\bibinfo{year}{2016}).
\newblock


\bibitem[\protect\citeauthoryear{Acun, Gupta, Jain, Langer, Menon, Mikida, Ni,
  Robson, Sun, Totoni, Wesolowski, and Kale}{Acun et~al\mbox{.}}{2014}]%
        {sc14charm}
\bibfield{author}{\bibinfo{person}{Bilge Acun}, \bibinfo{person}{Abhishek
  Gupta}, \bibinfo{person}{Nikhil Jain}, \bibinfo{person}{Akhil Langer},
  \bibinfo{person}{Harshitha Menon}, \bibinfo{person}{Eric Mikida},
  \bibinfo{person}{Xiang Ni}, \bibinfo{person}{Michael Robson},
  \bibinfo{person}{Yanhua Sun}, \bibinfo{person}{Ehsan Totoni},
  \bibinfo{person}{Lukasz Wesolowski}, {and} \bibinfo{person}{Laxmikant Kale}.}
  \bibinfo{year}{2014}\natexlab{}.
\newblock \showarticletitle{{Parallel Programming with Migratable Objects:
  Charm++ in Practice}} {\em (\bibinfo{series}{SC'14})}.
\newblock


\bibitem[\protect\citeauthoryear{Bishop}{Bishop}{2006}]%
        {Bishop:2006:PRM}
\bibfield{author}{\bibinfo{person}{Christopher~M. Bishop}.}
  \bibinfo{year}{2006}\natexlab{}.
\newblock \bibinfo{booktitle}{{\em Pattern Recognition and Machine Learning
  (Information Science and Statistics)}}.
\newblock \bibinfo{publisher}{Springer-Verlag New York, Inc.}
\newblock
\showISBNx{0387310738}


\bibitem[\protect\citeauthoryear{Blume, Doallo, Eigenmann, Grout, Hoeflinger,
  Lawrence, Lee, Padua, Paek, Pottenger, Rauchwerger, and Tu}{Blume
  et~al\mbox{.}}{1996}]%
        {Blume:1996:PPP}
\bibfield{author}{\bibinfo{person}{William Blume}, \bibinfo{person}{Ramon
  Doallo}, \bibinfo{person}{Rudolf Eigenmann}, \bibinfo{person}{John Grout},
  \bibinfo{person}{Jay Hoeflinger}, \bibinfo{person}{Thomas Lawrence},
  \bibinfo{person}{Jaejin Lee}, \bibinfo{person}{David Padua},
  \bibinfo{person}{Yunheung Paek}, \bibinfo{person}{Bill Pottenger},
  \bibinfo{person}{Lawrence Rauchwerger}, {and} \bibinfo{person}{Peng Tu}.}
  \bibinfo{year}{1996}\natexlab{}.
\newblock \showarticletitle{Parallel Programming with Polaris}.
\newblock \bibinfo{journal}{{\em Computer\/}} \bibinfo{volume}{29},
  \bibinfo{number}{12} (\bibinfo{year}{1996}).
\newblock


\bibitem[\protect\citeauthoryear{Bottou}{Bottou}{2010}]%
        {bottou2010large}
\bibfield{author}{\bibinfo{person}{L{\'e}on Bottou}.}
  \bibinfo{year}{2010}\natexlab{}.
\newblock \showarticletitle{Large-scale machine learning with stochastic
  gradient descent}.
\newblock In \bibinfo{booktitle}{{\em Proceedings of COMPSTAT'2010}}.
\newblock


\bibitem[\protect\citeauthoryear{Brown, Lee, Rompf, Sujeeth, De~Sa, Aberger,
  and Olukotun}{Brown et~al\mbox{.}}{2016}]%
        {brown2016have}
\bibfield{author}{\bibinfo{person}{Kevin~J. Brown}, \bibinfo{person}{HyoukJoong
  Lee}, \bibinfo{person}{Tiark Rompf}, \bibinfo{person}{Arvind~K. Sujeeth},
  \bibinfo{person}{Christopher De~Sa}, \bibinfo{person}{Christopher Aberger},
  {and} \bibinfo{person}{Kunle Olukotun}.} \bibinfo{year}{2016}\natexlab{}.
\newblock \showarticletitle{Have Abstraction and Eat Performance, Too:
  Optimized Heterogeneous Computing with Parallel Patterns}. In
  \bibinfo{booktitle}{{\em CGO}}.
\newblock


\bibitem[\protect\citeauthoryear{Catanzaro, Garland, and Keutzer}{Catanzaro
  et~al\mbox{.}}{2011}]%
        {Catanzaro:2011:CCE}
\bibfield{author}{\bibinfo{person}{Bryan Catanzaro}, \bibinfo{person}{Michael
  Garland}, {and} \bibinfo{person}{Kurt Keutzer}.}
  \bibinfo{year}{2011}\natexlab{}.
\newblock \showarticletitle{Copperhead: Compiling an Embedded Data Parallel
  Language} {\em (\bibinfo{series}{PPoPP '11})}.
\newblock


\bibitem[\protect\citeauthoryear{Chatterjee, Gilbert, Schreiber, and
  Teng}{Chatterjee et~al\mbox{.}}{1995}]%
        {Chatterjee:1995:OEA}
\bibfield{author}{\bibinfo{person}{Siddhartha Chatterjee},
  \bibinfo{person}{John~R. Gilbert}, \bibinfo{person}{Robert Schreiber}, {and}
  \bibinfo{person}{Shang-Hua Teng}.} \bibinfo{year}{1995}\natexlab{}.
\newblock \showarticletitle{Optimal Evaluation of Array Expressions on
  Massively Parallel Machines}.
\newblock \bibinfo{journal}{{\em ACM Trans. Program. Lang. Syst.\/}}
  \bibinfo{volume}{17}, \bibinfo{number}{1} (\bibinfo{year}{1995}).
\newblock


\bibitem[\protect\citeauthoryear{Chaves, Nehrbass, Guilfoos, Gardiner, Ahalt,
  Krishnamurthy, Unpingco, Chalker, Warnock, and Samsi}{Chaves
  et~al\mbox{.}}{2006}]%
        {Chaves06}
\bibfield{author}{\bibinfo{person}{J.~C. Chaves}, \bibinfo{person}{J.
  Nehrbass}, \bibinfo{person}{B. Guilfoos}, \bibinfo{person}{J. Gardiner},
  \bibinfo{person}{S. Ahalt}, \bibinfo{person}{A. Krishnamurthy},
  \bibinfo{person}{J. Unpingco}, \bibinfo{person}{A. Chalker},
  \bibinfo{person}{A. Warnock}, {and} \bibinfo{person}{S. Samsi}.}
  \bibinfo{year}{2006}\natexlab{}.
\newblock \showarticletitle{Octave and Python: High-Level Scripting Languages
  Productivity and Performance Evaluation}. In \bibinfo{booktitle}{{\em HPCMP
  Users Group Conference, 2006}}.
\newblock


\bibitem[\protect\citeauthoryear{Dean and Ghemawat}{Dean and Ghemawat}{2008}]%
        {dean2008mapreduce}
\bibfield{author}{\bibinfo{person}{Jeffrey Dean} {and} \bibinfo{person}{Sanjay
  Ghemawat}.} \bibinfo{year}{2008}\natexlab{}.
\newblock \showarticletitle{{M}ap{R}educe: simplified data processing on large
  clusters}.
\newblock \bibinfo{journal}{{\it Commun. ACM}} (\bibinfo{year}{2008}).
\newblock


\bibitem[\protect\citeauthoryear{Denniston, Kamil, and Amarasinghe}{Denniston
  et~al\mbox{.}}{2016}]%
        {denniston2016distributed}
\bibfield{author}{\bibinfo{person}{Tyler Denniston}, \bibinfo{person}{Shoaib
  Kamil}, {and} \bibinfo{person}{Saman Amarasinghe}.}
  \bibinfo{year}{2016}\natexlab{}.
\newblock \showarticletitle{Distributed Halide} {\em (\bibinfo{series}{PPoPP
  '16})}.
\newblock


\bibitem[\protect\citeauthoryear{Eigenmann, Hoeflinger, and Padua}{Eigenmann
  et~al\mbox{.}}{1998}]%
        {Eigenmann94onthe}
\bibfield{author}{\bibinfo{person}{R. Eigenmann}, \bibinfo{person}{J.
  Hoeflinger}, {and} \bibinfo{person}{D. Padua}.}
  \bibinfo{year}{1998}\natexlab{}.
\newblock \showarticletitle{On the automatic parallelization of the Perfect
  Benchmarks(R)}.
\newblock \bibinfo{journal}{{\em IEEE Transactions on Parallel and Distributed
  Systems\/}} \bibinfo{volume}{9}, \bibinfo{number}{1} (\bibinfo{year}{1998}).
\newblock


\bibitem[\protect\citeauthoryear{Hiranandani, Kennedy, Tseng, and
  Warren}{Hiranandani et~al\mbox{.}}{1994}]%
        {Hiranandani:1994:DEN}
\bibfield{author}{\bibinfo{person}{Seema Hiranandani}, \bibinfo{person}{Ken
  Kennedy}, \bibinfo{person}{Chau~Wen Tseng}, {and} \bibinfo{person}{Scott
  Warren}.} \bibinfo{year}{1994}\natexlab{}.
\newblock \showarticletitle{The D Editor: A New Interactive Parallel
  Programming Tool} {\em (\bibinfo{series}{Supercomputing '94})}.
\newblock


\bibitem[\protect\citeauthoryear{Islam, Rahman, Lu, Shankar, and Panda}{Islam
  et~al\mbox{.}}{2015}]%
        {clusterApache15}
\bibfield{author}{\bibinfo{person}{N. Islam}, \bibinfo{person}{W. Rahman},
  \bibinfo{person}{X. Lu}, \bibinfo{person}{D. Shankar}, {and}
  \bibinfo{person}{D. Panda}.} \bibinfo{year}{2015}\natexlab{}.
\newblock \showarticletitle{Performance Characterization and Acceleration of
  In-Memory File Systems for Hadoop and Spark Applications on HPC Clusters}. In
  \bibinfo{booktitle}{{\em 2015 IEEE International Conference on Big Data}}.
\newblock


\bibitem[\protect\citeauthoryear{Kaiser, Heller, Adelstein-Lelbach, Serio, and
  Fey}{Kaiser et~al\mbox{.}}{2014}]%
        {Kaiser:2014:HPX}
\bibfield{author}{\bibinfo{person}{Hartmut Kaiser}, \bibinfo{person}{Thomas
  Heller}, \bibinfo{person}{Bryce Adelstein-Lelbach}, \bibinfo{person}{Adrian
  Serio}, {and} \bibinfo{person}{Dietmar Fey}.}
  \bibinfo{year}{2014}\natexlab{}.
\newblock \showarticletitle{{HPX}: A Task Based Programming Model in a Global
  Address Space} {\em (\bibinfo{series}{PGAS '14})}.
\newblock


\bibitem[\protect\citeauthoryear{Kennedy, Koelbel, and Zima}{Kennedy
  et~al\mbox{.}}{2007}]%
        {Kennedy:2007:RFH}
\bibfield{author}{\bibinfo{person}{Ken Kennedy}, \bibinfo{person}{Charles
  Koelbel}, {and} \bibinfo{person}{Hans Zima}.}
  \bibinfo{year}{2007}\natexlab{}.
\newblock \showarticletitle{The Rise and Fall of High Performance Fortran: An
  Historical Object Lesson} {\em (\bibinfo{series}{HOPL III})}.
\newblock


\bibitem[\protect\citeauthoryear{Kennedy and Kremer}{Kennedy and
  Kremer}{1998}]%
        {Kennedy:1998:ADL}
\bibfield{author}{\bibinfo{person}{Ken Kennedy} {and} \bibinfo{person}{Ulrich
  Kremer}.} \bibinfo{year}{1998}\natexlab{}.
\newblock \showarticletitle{Automatic Data Layout for Distributed-memory
  Machines}.
\newblock \bibinfo{journal}{{\em ACM Trans. Program. Lang. Syst.\/}}
  \bibinfo{volume}{20}, \bibinfo{number}{4} (\bibinfo{year}{1998}).
\newblock
\showISSN{0164-0925}


\bibitem[\protect\citeauthoryear{Lion, Chiu, Sun, Zhuang, Grcevski, and
  Yuan}{Lion et~al\mbox{.}}{2016}]%
        {david2016}
\bibfield{author}{\bibinfo{person}{David Lion}, \bibinfo{person}{Adrian Chiu},
  \bibinfo{person}{Hailong Sun}, \bibinfo{person}{Xin Zhuang},
  \bibinfo{person}{Nikola Grcevski}, {and} \bibinfo{person}{Ding Yuan}.}
  \bibinfo{year}{2016}\natexlab{}.
\newblock \showarticletitle{Don{\textquoteright}t Get Caught in the Cold,
  Warm-up Your {JVM}: Understand and Eliminate {JVM} Warm-up Overhead in
  Data-Parallel Systems}. In \bibinfo{booktitle}{{\em OSDI}}.
  \bibinfo{publisher}{USENIX Association}.
\newblock


\bibitem[\protect\citeauthoryear{Maas, Harris, Asanovi{\'c}, and
  Kubiatowicz}{Maas et~al\mbox{.}}{2015}]%
        {martin2015}
\bibfield{author}{\bibinfo{person}{Martin Maas}, \bibinfo{person}{Tim Harris},
  \bibinfo{person}{Krste Asanovi{\'c}}, {and} \bibinfo{person}{John
  Kubiatowicz}.} \bibinfo{year}{2015}\natexlab{}.
\newblock \showarticletitle{Trash Day: Coordinating Garbage Collection in
  Distributed Systems}. In \bibinfo{booktitle}{{\em {HotOS}}}.
\newblock


\bibitem[\protect\citeauthoryear{McSherry, Isard, and Murray}{McSherry
  et~al\mbox{.}}{2015}]%
        {mcsherry2015scalability}
\bibfield{author}{\bibinfo{person}{Frank McSherry}, \bibinfo{person}{Michael
  Isard}, {and} \bibinfo{person}{Derek~G Murray}.}
  \bibinfo{year}{2015}\natexlab{}.
\newblock \showarticletitle{Scalability! but at what COST?} {\em
  (\bibinfo{series}{HotOS})}.
\newblock


\bibitem[\protect\citeauthoryear{Newburn, So, Liu, McCool, Ghuloum, Toit, Wang,
  Du, Chen, Wu, Guo, Liu, and Zhang}{Newburn et~al\mbox{.}}{2011}]%
        {Newburn:2011:IAB}
\bibfield{author}{\bibinfo{person}{Chris~J. Newburn}, \bibinfo{person}{Byoungro
  So}, \bibinfo{person}{Zhenying Liu}, \bibinfo{person}{Michael McCool},
  \bibinfo{person}{Anwar Ghuloum}, \bibinfo{person}{Stefanus~Du Toit},
  \bibinfo{person}{Zhi~Gang Wang}, \bibinfo{person}{Zhao~Hui Du},
  \bibinfo{person}{Yongjian Chen}, \bibinfo{person}{Gansha Wu},
  \bibinfo{person}{Peng Guo}, \bibinfo{person}{Zhanglin Liu}, {and}
  \bibinfo{person}{Dan Zhang}.} \bibinfo{year}{2011}\natexlab{}.
\newblock \showarticletitle{Intel's Array Building Blocks: A Retargetable,
  Dynamic Compiler and Embedded Language} {\em (\bibinfo{series}{CGO '11})}.
\newblock


\bibitem[\protect\citeauthoryear{Nielson, Nielson, and Hankin}{Nielson
  et~al\mbox{.}}{2015}]%
        {nielson2015principles}
\bibfield{author}{\bibinfo{person}{Flemming Nielson}, \bibinfo{person}{Hanne~R
  Nielson}, {and} \bibinfo{person}{Chris Hankin}.}
  \bibinfo{year}{2015}\natexlab{}.
\newblock \bibinfo{booktitle}{{\em Principles of program analysis}}.
\newblock \bibinfo{publisher}{Springer}.
\newblock


\bibitem[\protect\citeauthoryear{Ousterhout, Rasti, Ratnasamy, Shenker, and
  Chun}{Ousterhout et~al\mbox{.}}{2015}]%
        {OusterhoutAnalysis15}
\bibfield{author}{\bibinfo{person}{Kay Ousterhout}, \bibinfo{person}{Ryan
  Rasti}, \bibinfo{person}{Sylvia Ratnasamy}, \bibinfo{person}{Scott Shenker},
  {and} \bibinfo{person}{Byung-Gon Chun}.} \bibinfo{year}{2015}\natexlab{}.
\newblock \showarticletitle{Making Sense of Performance in Data Analytics
  Frameworks} {\em (\bibinfo{series}{NSDI'15})}.
\newblock


\bibitem[\protect\citeauthoryear{Plank, Beck, and Kingsley}{Plank
  et~al\mbox{.}}{1995}]%
        {pbk:95:came}
\bibfield{author}{\bibinfo{person}{J.~S. Plank}, \bibinfo{person}{M. Beck},
  {and} \bibinfo{person}{G. Kingsley}.} \bibinfo{year}{1995}\natexlab{}.
\newblock \showarticletitle{Compiler-Assisted Memory Exclusion for Fast
  Checkpointing}.
\newblock \bibinfo{journal}{{\em IEEE Technical Committee on Operating Systems
  and Application Environments\/}} \bibinfo{volume}{7}, \bibinfo{number}{4}
  (\bibinfo{year}{1995}).
\newblock


\bibitem[\protect\citeauthoryear{Prechelt}{Prechelt}{2000}]%
        {Prechelt:2000}
\bibfield{author}{\bibinfo{person}{Lutz Prechelt}.}
  \bibinfo{year}{2000}\natexlab{}.
\newblock \showarticletitle{An Empirical Comparison of Seven Programming
  Languages}.
\newblock \bibinfo{journal}{{\em IEEE Computer\/}} \bibinfo{volume}{33},
  \bibinfo{number}{10} (\bibinfo{date}{Oct.} \bibinfo{year}{2000}),
  \bibinfo{pages}{23--29}.
\newblock


\bibitem[\protect\citeauthoryear{Reyes{-}Ortiz, Oneto, and
  Anguita}{Reyes{-}Ortiz et~al\mbox{.}}{2015}]%
        {ReyesOrtizMPIspark2015}
\bibfield{author}{\bibinfo{person}{Jorge~Luis Reyes{-}Ortiz},
  \bibinfo{person}{Luca Oneto}, {and} \bibinfo{person}{Davide Anguita}.}
  \bibinfo{year}{2015}\natexlab{}.
\newblock \showarticletitle{Big Data Analytics in the Cloud: Spark on Hadoop vs
  MPI/OpenMP on Beowulf}. In \bibinfo{booktitle}{{\em {INNS} Conference on Big
  Data 2015}}.
\newblock


\bibitem[\protect\citeauthoryear{Schulz, Bronevetsky, Fernandes, Marques,
  Pingali, and Stodghill}{Schulz et~al\mbox{.}}{2004}]%
        {Schulz:2004:IES}
\bibfield{author}{\bibinfo{person}{Martin Schulz}, \bibinfo{person}{Greg
  Bronevetsky}, \bibinfo{person}{Rohit Fernandes}, \bibinfo{person}{Daniel
  Marques}, \bibinfo{person}{Keshav Pingali}, {and} \bibinfo{person}{Paul
  Stodghill}.} \bibinfo{year}{2004}\natexlab{}.
\newblock \showarticletitle{Implementation and Evaluation of a Scalable
  Application-Level Checkpoint-Recovery Scheme for MPI Programs} {\em
  (\bibinfo{series}{SC '04})}.
\newblock


\bibitem[\protect\citeauthoryear{Sujeeth, Brown, Lee, Rompf, Chafi, Odersky,
  and Olukotun}{Sujeeth et~al\mbox{.}}{2014}]%
        {Sujeeth:2014:DCA}
\bibfield{author}{\bibinfo{person}{Arvind~K. Sujeeth},
  \bibinfo{person}{Kevin~J. Brown}, \bibinfo{person}{Hyoukjoong Lee},
  \bibinfo{person}{Tiark Rompf}, \bibinfo{person}{Hassan Chafi},
  \bibinfo{person}{Martin Odersky}, {and} \bibinfo{person}{Kunle Olukotun}.}
  \bibinfo{year}{2014}\natexlab{}.
\newblock \showarticletitle{Delite: A Compiler Architecture for
  Performance-Oriented Embedded Domain-Specific Languages}.
\newblock \bibinfo{journal}{{\em ACM Trans. Embed. Comput. Syst.\/}}
  \bibinfo{volume}{13}, \bibinfo{number}{4s} (\bibinfo{year}{2014}).
\newblock


\bibitem[\protect\citeauthoryear{Waheed, Frumkin, Yan, Jin, and Hribar}{Waheed
  et~al\mbox{.}}{1998}]%
        {Waheed99cap}
\bibfield{author}{\bibinfo{person}{Abdul Waheed}, \bibinfo{person}{Michael
  Frumkin}, \bibinfo{person}{Jerry Yan}, \bibinfo{person}{Haoqiang Jin}, {and}
  \bibinfo{person}{Michelle Hribar}.} \bibinfo{year}{1998}\natexlab{}.
\newblock \showarticletitle{A Comparison of Automatic Parallelization
  Tools/Compilers on the SGI Origin 2000}.
\newblock \bibinfo{journal}{{\em SC'98\/}} (\bibinfo{year}{1998}).
\newblock


\bibitem[\protect\citeauthoryear{Wahlberg, Boyd, Annergren, and Wang}{Wahlberg
  et~al\mbox{.}}{2012}]%
        {wahlberg2012admm}
\bibfield{author}{\bibinfo{person}{Bo Wahlberg}, \bibinfo{person}{Stephen
  Boyd}, \bibinfo{person}{Mariette Annergren}, {and} \bibinfo{person}{Yang
  Wang}.} \bibinfo{year}{2012}\natexlab{}.
\newblock \showarticletitle{An ADMM algorithm for a class of total variation
  regularized estimation problems}.
\newblock \bibinfo{journal}{{\em IFAC Proceedings Volumes\/}}
  \bibinfo{volume}{45}, \bibinfo{number}{16} (\bibinfo{year}{2012}),
  \bibinfo{pages}{83--88}.
\newblock


\bibitem[\protect\citeauthoryear{Wang, Yang, Fu, Du, Wang, and Jia}{Wang
  et~al\mbox{.}}{2008}]%
        {Wang:2008:AAC}
\bibfield{author}{\bibinfo{person}{Panfeng Wang}, \bibinfo{person}{Xuejun
  Yang}, \bibinfo{person}{Hongyi Fu}, \bibinfo{person}{Yunfei Du},
  \bibinfo{person}{Zhiyun Wang}, {and} \bibinfo{person}{Jia Jia}.}
  \bibinfo{year}{2008}\natexlab{}.
\newblock \showarticletitle{Automated Application-level Checkpointing Based on
  Live-variable Analysis in MPI Programs} {\em (\bibinfo{series}{PPoPP '08})}.
  \bibinfo{pages}{273--274}.
\newblock


\bibitem[\protect\citeauthoryear{White}{White}{2012}]%
        {white2012hadoop}
\bibfield{author}{\bibinfo{person}{Tom White}.}
  \bibinfo{year}{2012}\natexlab{}.
\newblock \bibinfo{booktitle}{{\em Hadoop: The definitive guide}}.
\newblock \bibinfo{publisher}{{O'Reilly Media, Inc.}}
\newblock


\bibitem[\protect\citeauthoryear{Wilson, French, Wilson, Amarasinghe, Anderson,
  Tjiang, Liao, Tseng, Hall, Lam, and Hennessy}{Wilson et~al\mbox{.}}{1994}]%
        {Wilson:1994:SIR}
\bibfield{author}{\bibinfo{person}{Robert~P. Wilson},
  \bibinfo{person}{Robert~S. French}, \bibinfo{person}{Christopher~S. Wilson},
  \bibinfo{person}{Saman~P. Amarasinghe}, \bibinfo{person}{Jennifer~M.
  Anderson}, \bibinfo{person}{Steve W.~K. Tjiang}, \bibinfo{person}{Shih-Wei
  Liao}, \bibinfo{person}{Chau-Wen Tseng}, \bibinfo{person}{Mary~W. Hall},
  \bibinfo{person}{Monica~S. Lam}, {and} \bibinfo{person}{John~L. Hennessy}.}
  \bibinfo{year}{1994}\natexlab{}.
\newblock \showarticletitle{SUIF: An Infrastructure for Research on
  Parallelizing and Optimizing Compilers}.
\newblock \bibinfo{journal}{{\em SIGPLAN Not.\/}} \bibinfo{volume}{29},
  \bibinfo{number}{12} (\bibinfo{year}{1994}).
\newblock


\bibitem[\protect\citeauthoryear{Young}{Young}{1974}]%
        {Young}
\bibfield{author}{\bibinfo{person}{John~W. Young}.}
  \bibinfo{year}{1974}\natexlab{}.
\newblock \showarticletitle{A First Order Approximation to the Optimum
  Checkpoint Interval}.
\newblock \bibinfo{journal}{{\em Commun. ACM\/}} \bibinfo{volume}{17},
  \bibinfo{number}{9} (\bibinfo{date}{Sept.} \bibinfo{year}{1974}),
  \bibinfo{pages}{530--531}.
\newblock


\bibitem[\protect\citeauthoryear{Zaharia, Chowdhury, Franklin, Shenker, and
  Stoica}{Zaharia et~al\mbox{.}}{2010}]%
        {zaharia2010spark}
\bibfield{author}{\bibinfo{person}{Matei Zaharia}, \bibinfo{person}{Mosharaf
  Chowdhury}, \bibinfo{person}{Michael~J. Franklin}, \bibinfo{person}{Scott
  Shenker}, {and} \bibinfo{person}{Ion Stoica}.}
  \bibinfo{year}{2010}\natexlab{}.
\newblock \showarticletitle{{Spark}: Cluster Computing with Working Sets}.
\newblock  (\bibinfo{year}{2010}).
\newblock


\bibitem[\protect\citeauthoryear{Zhang, Ruan, and Qiu}{Zhang
  et~al\mbox{.}}{2015}]%
        {zhang2015harp}
\bibfield{author}{\bibinfo{person}{Bingjing Zhang}, \bibinfo{person}{Yang
  Ruan}, {and} \bibinfo{person}{Judy Qiu}.} \bibinfo{year}{2015}\natexlab{}.
\newblock \showarticletitle{Harp: Collective communication on hadoop} {\em
  (\bibinfo{series}{IC2E})}.
\newblock


\end{thebibliography}

\end{document}